\documentclass[12pt]{iopart}
\usepackage{epsfig}
\begin{document}

\title[]{Stability and electronic structure in hexagonal
$\rm \beta\,Al_9Mn_3Si$ and $\rm \varphi\,Al_{10}Mn_3$ crystals}

\author{Guy Trambly de Laissardi\`ere}

\address{Laboratoire de Physique Th\'eorique et Mod\'elisation,
CNRS--Universit\'e de Cergy-Pontoise (ESA 8089),
5 mail Gay-Lussac, Neuville sur Oise,
95031 Cergy-Pontoise, France}

\ead{guy.trambly@ptm.u-cergy.fr}

\begin{abstract}
The electronic structures of hexagonal $\rm
\beta\,Al_9Mn_3Si$ and $\rm \varphi\,Al_{10}Mn_3$ 
are investigated through
self-consistent calculations carried out using the LMTO method. This
{\it ab initio} approach is combined with an analysis of a simplified
hamiltonian model for Al based alloys containing transition metal
atoms. 
Results show a strong effect on an atomic structure stabilisation
by an indirect Mn-Mn
interaction mediated by conduction electrons over 
medium range distances
($\rm 5\,\AA$ and more). 
Both the role and position of Si atoms are explained
as well as  the origin of 
large vacancies which characterises these atomic structures. 
As $\beta$
and $\varphi$ phases are related to Al based quasicrystals and 
related
approximant structures, 
it yields arguments on the stabilisation of
such complex phases.

\end{abstract}

\submitto{\JPCM}
\hspace{12cm} 2002/7/5
\maketitle

\section{Introduction}
\label{SecIntro}

The almost isomorphic stable $\rm \beta\,Al_9Mn_3Si$
\cite{Robinson52}
and metastable $\rm \varphi\,Al_{10}Mn_3$
\cite{Taylor59}
phases
are often present in alloys containing quasicrystals
in Al(Si)-Mn systems.
Although their diffraction features  are
different from those of quasicrystals,
several correlations with quasiperiodic atomic structure
have been shown.
For instance, there is a strong resemblance of these phases
with parts of the complex structures of $\rm \mu\,Al_{4.12}Mn$
\cite{Shoemaker89}
and $\rm \lambda\,Al_4Mn$ \cite{Kreiner97} which are
related to quasicrystals.
$\beta$
and  $\varphi$ are also almost isomorphic with
$\rm Al_5Co_2$ \cite{Newkirk61} which is an approximant
of decagonal quasicrystal with the shortest periodic
stacking sequence along the tenfold axis \cite{Song92}.

Meta-stable
icosahedral ($i$-) and decagonal
($d$-) quasicrystals have been found in the  Al-Mn system 
\cite{Shechtman84,Berger94,Harmelin94}.
With the addition of few per cent of Si atoms,
new stable phases are obtained:
$i$-Al-Mn-Si \cite{Berger94},
approximant
$\rm \alpha\,Al_9Mn_2Si$
\cite{Cooper66,Guyot85,ElserH85}
and $\rm \beta\,Al_9Mn_3Si$\dots
The occurrence of stable complex structure in Al-Mn and Al-Mn-Si
is a major question in the understanding of the stability
of quasicrystals.
For instance, the role of Si in stabilising the
$i$-phase, is not yet understood.
In this direction, investigations on relations between
isomorphic stable $\rm \beta\,Al_9Mn_3Si$
and meta-stable $\rm \varphi\,Al_{10}Mn_3$  phases represent a great
interest.
On another hand, these phases give a good example 
to analyse the 
effect of the position of transition metal (TM) 
atoms in stabilising complex structure related to 
quasiperiodicity.

In this paper, a first-principles 
({\it ab initio}) study of the
electronic structure in $\rm \beta\,Al_9Mn_3Si$ and
$\rm \varphi\,Al_{10}Mn_3$ phases 
is combined with a model
approach
in order to describe the interplay between the medium range
order and the electronic structure. 
Results are compared between $\beta$, $\varphi$, $\rm Al_5Co_2$, 
$\rm \mu\,Al_{4.12}Mn$,
and $\rm \lambda\,Al_4Mn$ phases.
The stabilising role of Si  and
the origin of a large hole (vacancy)
in both $\beta$ 
and $\varphi$ phases are justified.
The origin of a pseudogap is analysed in the frame  
of 
the Hume-Rothery stabilisation rule 
for sp-d electron phases
\cite{GuyEuroPhys93,GuyPRB95}.
Besides,
a  real space approach in term of
a realistic
TM-TM pair interaction allows
to understand the effect of Mn position.
As these phases
are related to quasiperiodic phases,
such a study yields 
arguments to discuss the interplay
between electronic structure and stability in quasicrystals.

The paper is organised as follows.
In section \ref{Sec_HR}, 
is presented a short review on Hume-Rothery mechanism
in
Al(rich)-TM phases that
has often been proposed for the stabilisation of 
crystals and quasicrystals.
The structures of $\varphi$ and
$\beta$ are presented in section \ref{SecStructure}
with a discussion on their relations with
quasicrystals.
First-principles ({\it ab initio})
study of the
electronic structure is presented
in section \ref{SecLMTO}.
Then the effect of the sp-d hybridisation is analysed in details
through {\it ab initio} calculations for hypothetical structures.
In section \ref{BraggPotential}, 
these results are understood in term of a
Friedel-Anderson sp-d hamiltonian that allows
to find the {\it ``effective Bragg potential''} 
for sp-d Hume-Rothery
alloys.
In section \ref{SecMn_MnInteraction},
a real space approach of the Hume-Rothery
mechanism shows the strong effect of a medium range
Mn-Mn pair interaction (up to $\sim5\,\rm \AA$ and more).
Magnetism is studied in  section \ref{SecMagnetism}
and a short conclusion is given in section 
\ref{SecConclusion}.

%
%
\section{Hume-Rothery stabilisation in quasicrystals 
and related crystals}
\label{Sec_HR}

\subsection{Near contact between Fermi sphere and pseudo-Brillouin zone}

Since the 1950s, Al(rich)-TM crystals
are considered by many authors as Hume-Rothery alloys
\cite{HumeRothery54}
(for instance see Refs
\cite{Raynor49,Douglas49,Nicol53,
Raynor49b,HumeRothery54b,Robinson52,Taylor59}).
In these phases, the important parameter is the average
number of electrons
per atom, $e/a$.
The valence of Al and Si are fixed without ambiguity (+3 and
+4, respectively).
Following classical theory 
\cite{Raynor49b,HumeRothery54b}, a negative valence is
assigned to TM atom (typically, $-3$ for Mn,
$-2$ for Fe, $-1$ for Co and $0$ for Ni).
For
$\rm \varphi\,Al_{10}Mn_3$,
$\rm \beta\,Al_9Mn_3Si$ and
$\rm Al_5Co_2$,
$e/a$ is equal to 1.61, 1.69 and 1.86, respectively.
The occurrence of different compounds with similar structures
is therefore to be explained by the fact
that they are electron compounds with
similar e/a ratio in spite of different atomic
concentrations \cite{Robinson52}.
Indeed, for these phases a
band energy minimisation occurs when the Fermi sphere touches
a pseudo-Brillouin zone (prominent Brillouin zone),
constructed by Bragg vectors
${\bf K}_p$ corresponding to intense peaks in the experimental
diffraction pattern.
The Hume-Rothery condition for alloying
is then $2k_F\simeq K_p$.
Assuming a free electron valence
band, the Fermi momentum, $k_F$, is calculated
from $e/a$.

Soon after the discovery of quasicrystals,
it has been pointed out that their stoechiometry 
appears to be
governed by a Hume-Rothery rule
(see for instance
\cite{Friedel87,Friedel88,Smith87,FujiAlMnSi,
Tsai91,Fujiwara91,Hafner92,Mayou94}).
Indeed the $e/a$ ratio has been used for a long time
to distinguish between Frank-Kasper type quasicrystals
(sp quasicrystals)
and Mackay type quasicrystals (sp-d quasicrystals) 
\cite{Tsai91}.
Friedel and D\'enoyer \cite{Friedel87} 
have determined 
the pseudo-Brillouin zone
in contact with the Fermi sphere for $i$-Al-Li-Cu.
Gratias et al. \cite{Gratias93} have shown
that the Al-Cu-Fe icosahedral domain is
located along a line in the phase diagram defined by the
equation $\rm e/a \simeq 1.86$.
Besides,
Al-Cu-Fe alloys along an $e/a$-constant
line have similar local electronic properties and local atomic
order \cite{Hippert94}.
Recently, a pseudo-Brillouin zone that touches the
Fermi Sphere in 1/1\,Al-Cu-Ru-Si and Al-Mg-Zn approximants
has been identified \cite{Sato01PRB,MizutaniMRS01}.
With the discovery of decagonal 
$d$-Al-Cu-Co and $d$-Al-Ni-Co by
Tsai et al. \cite{Tsai89},
these authors \cite{Tsai91} have determined 
that the value of
$e/a$ ratio is about 1.7
in spite of wide composition range for
quasicrystals in these systems.
The importance of $e/a$ value in quasicrystals and their
properties suggests that Hume-Rothery
mechanism plays a significant role in their stabilisation.

\subsection{Pseudogap in the density of states}

The density of states (DOS) in sp Hume-Rothery alloys is well
described by the {\it Jones theory}
(for review see Refs 
\cite{Massalski78,Paxton97,Pettifor00}).
The valence band (sp states) are nearly-free electrons, and
the Fermi-sphere\,/\,pseudo-Brillouin zone
interaction creates a depletion in the DOS,
called ``{\it pseudogap}'',
near the Fermi energy, $\rm E_F$.
This pseudogap has been found both experimentally and
from first-principles calculations in classical Hume-Rothery
alloys (see for instance the recent theoretical study of 
archetypal system Cu-Zn \cite{Paxton97}).
It has also been found 
experimentally and theoretically in sp quasicrystals 
and related phases
(for instance in Al-Li-Cu
\cite{Berger94,Fujiwara91}
and Al-Mg-Zn \cite{Hafner92,Sato01PRB}).
But, the treatment of
Al(rich) alloys containing transition metal elements
requires a new theory.
Indeed, the d states of TM are not nearly-free states in spite
of strong sp-d hybridisation.
Thus a model for sp-d electron phases which combined the
effect of the diffraction by Bragg planes with the
sp-d hybridisation has been developped
\cite{GuyEuroPhys93,GuyPRB95}.
It is shown that negative valence of TM atom
results from particular effects of the sp-d hybridisation
in Hume-Rothery alloys
\cite{Friedel88,Friedel92,GuyEuroPhys93,GuyPRB95}.
Besides the TM DOS (mainly d states) depends strongly
on TM atoms positions.
For particular TM
positions, one obtains a pseudogap near $E_F$ in total
DOS and partial d DOS.
This has been confirmed by {\it ab initio}
calculations in a series of Al(rich)-TM crystals including
$\rm Al_5Co_2$ \cite{GuyPRB95},which is isomorphic with 
$\rm \beta\,Al_9Mn_3Si$ and
$\rm \varphi\,Al_{10}Mn_3$.
The presence of a pseudogap in
$\rm Al_5Co_2$ DOS
has also been confirmed by
photoemission spectroscopy \cite{Belin97}.
For icosahedral sp-d quasicrystals and their approximants, 
a wide
pseudogap at $\rm E_F$ has been found
experimentally \cite{Berger94,Mori91,Belin92,Belin94a,Belin94b,
Stadnik01,Fournee02}
and from {\it ab initio} calculations
\cite{FujiAlMnSi,GuyPRB94_AlCuFe,Krajvci95,Hafner98a,MizutaniMRS01}.
For instance in
$i$-Al-Cu-Fe, $i$-Al-Pd-Mn
and $\alpha$\,Al-Mn-Si, the DOS at $\rm E_F$ is reduced 
by $\sim 1/3$ with respect to pure Al (c.f.c.) DOS
\cite{Berger94}.

However, there are contradictory
results  about  DOS in decagonal
quasicrystals.
Photoemission spectroscopy measurements in the photon-energy range
35-120\,eV do not show any pseudogap \cite{Stadnik95}
in  $d$-$\rm Al_{65}Co_{15}Cu_{20}$ and
$d$-$\rm Al_{70}Co_{15}Ni_{15}$, whereas
ultrahigh resolution ultraviolet photoemission
shows a depletion of the DOS at $\rm E_F$
for the same compositions \cite{Stadnik97}.
From soft X-ray spectroscopy,
DOS in $d$-$\rm Al_{65}Co_{20}Cu_{15}$
and  $d$-$\rm Al_{70}Co_{15}Ni_{15}$, exhibits also a
pseudogap in the Al-3p band \cite{Belin96}.
Recently
a pseudogap, enhanced by sp-d hybridisation,
has been found in the Al-p band of $d$-Al-Pd-Mn
\cite{Fournee02}.
There are also {\it ab initio} calculations
performed for several atomic model approximants of
$d$-Al-Cu-Co
\cite{GuyPRBAlCuCo,Sabiryanov95},
$d$-Al-Co-Ni \cite{Krajci97_dAlNiCo}
and $d$-Al-Pd-Mn \cite{Krajci97_dAlPdMn}.
The results show that an existence of  pseudogap
depends on the position of the TM atoms.
Indeed, some TM atoms may ``fill up'' the
pseudogap, via the sp-d hybridisation;
whereas other TM positions enhance the pseudogap.

In summary, the importance of Hume-Rothery mechanism is now
established for many Al-based quasicrystals
with and without TM elements although
the presence of a pseudogap near $E_F$
is still discussed
for decagonal phases.
Nevertheless, one can 
not ignore the possible Hume-Rothery stabilising effect
on the 
origin of the quasiperiodicity.
This is the reason why in this
article {\it ab initio} results are analysed
in the
frame work of Hume-Rothery mechanism in order
to test the importance of this mechanism.

\section{Structures  and relations with quasicrystals}
\label{SecStructure}

\begin{table}
\caption{\label{Tab_Structure} Lattice parameters and
atomic positions of hexagonal 
$\rm \beta\,Al_9Mn_3Si$,
$\rm \varphi\,Al_{10}Mn_3$ and
$\rm Al_5Co_2$ phases of
$\rm P6_3/mmc$ space group. 
}
\begin{indented}
\item[]\begin{tabular}{@{}lllllllll}
\br
Lattice  &\multicolumn{2}{l}{$\rm \beta\,Al_9Mn_3Si$ \cite{Robinson52}} &
\multicolumn{2}{l}{$\rm \varphi\,Al_{10}Mn_3$ \cite{Taylor59}} &
\multicolumn{2}{l}{$\rm Al_5Co_2$ \cite{Newkirk61}}\\
parameters      &          &&          &&      &&         \\
$a$ (\AA)       & 7.513   &  & 7.543    & &  7.656 \\
$c$ (\AA)       & 7.745   &  & 7.898    & &  7.593 \\
\mr
Wyckoff    &          &&          &&      &&         \\
Sites      &          &&          &&      &&         \\
 (2a) : 0, 0, 0     & (Al,Si)(0)          && Al(0)           && Al(0)          \\
 (6h) : $x$,$2x$,$\frac{1}{4}$& (Al,Si)(1) & $x=.4579$ & Al(1) & $x=.4550$ & Al(1) & $x=.4702$\\
 (12k) : $x$,$2x$,$z$         & (Al,Si)(2) & $x=.2006$ & Al(2) & $x=.1995$ & Al(2) & $x=.1946$\\
                            &       & $z=-.0682$&       & $z=-.0630$&       & $z=-.0580$\\
 (6h) : $x$,$2x$,$\frac{1}{4}$& Mn    & $x=.1192$ & Mn    & $x=.1215$ & Co(1) & $x=.1268$\\
 (2d) : $\frac{2}{3}$,$\frac{1}{3}$,$\frac{1}{4}$& Va    &           & Va    &           & Co(0) &  \\
\br
\end{tabular}
\end{indented}
\end{table}

\begin{table}
\caption{\label{Tab_Distances} Interatomic distances
in $\rm \beta\,Al_9Mn_3Si$,
$\rm \varphi\,Al_{10}Mn_3$ and  $\rm Al_5Co_2$.
TM is either Mn or Co(1). 
$X$ corresponds to the vacancy  in
$\beta$ and $\varphi$ phases
and to Co(0) in $\rm Al_5Co_2$.
}
\begin{indented}
\item[]
\begin{tabular}{@{}llllll}
\br
Atom & Wyckoff      & Neighbours & Distances ($\rm \AA$) & & \\
     & site     &
& $\rm \beta\,Al_9Mn_3Si$ & $\rm \varphi\,Al_{10}Mn_3$
& $\rm Al_5Co_2$  \\
\mr
Al,Si(0) & (2a) & 6 Al(2) & 2.66 & 2.65 & 2.62 \\
         &      & 6 TM(1) & 2.48 & 2.53 & 2.53 \\
Al(1)  &(6h)    & 2 Al(1) & 2.81 & 2.75 & 3.14 \\
       &        & 4 Al(2) & 2.77 & 2.84 & 2.74 \\
       &        & 4 Al(2) & 2.98 & 2.99 & 2.97 \\
       &        & 2 TM(1) & 2.42 & 2.41 & 2.41 \\
       &        & 1 $X$     & 2.72 & 2.77 & 2.61 \\
Al(2)  &(12k)   &1 Al,Si(0)&2.66 & 2.65 & 2.62 \\
       &        & 2 Al(1) & 2.77 & 2.84 & 2.74 \\
       &        & 2 Al(1) & 2.98 & 2.99 & 2.97 \\
       &        & 2 Al(2) & 2.81 & 2.79 & 2.73 \\
       &        & 1 Al(2) & 2.82 & 2.95 & 2.92 \\
       &        & 2 Al(2) & 2.99 & 3.03 & 3.19 \\
       &        & 1 TM(1) & 2.68 & 2.67 & 2.51 \\
       &        & 2 TM(1) & 2.68 & 2.71 & 2.70 \\
       &        & 1 $X$     & 2.23 & 2.29 & 2.35 \\
TM(1)  &(6h)    &2 Al,Si(0)&2.48 & 2.53 & 2.54 \\
       &        & 2 Al(1) & 2.42 & 2.41 & 2.41 \\
       &        & 2 Al(2) & 2.68 & 2.67 & 2.51 \\
       &        & 4 Al(2) & 2.68 & 2.71 & 2.70 \\
       &        & 2 TM(1) & 2.69 & 2.75 & 2.91 \\
$X$ &(2d)       & 3 Al(1) & 2.72 & 2.77 & 2.61 \\
     &          & 6 Al(2) & 2.23 & 2.29 & 2.35 \\
     &          & 6 TM(1)$\rm ^{*}$& 3.81 & 3.82 & 3.86 \\
\br
\end{tabular}\\
$^{*}$ $X$ and TM(1) are not first-neighbour. \\
\end{indented}
\end{table}

\begin{figure}
\hspace{2cm}
\psfig{file=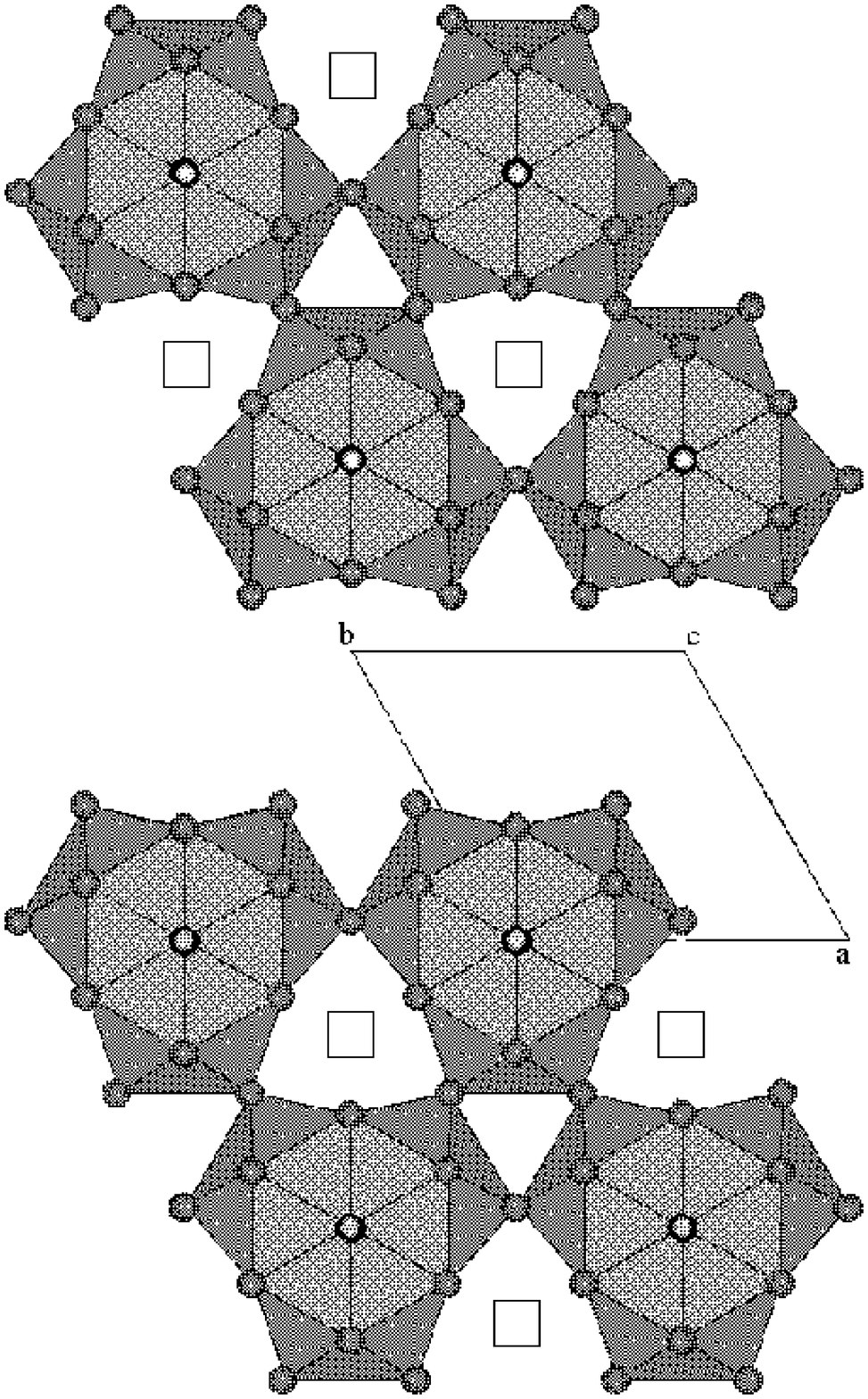,width=6cm}

\vspace{-9.5cm}
\hspace{9.5cm}
\psfig{file=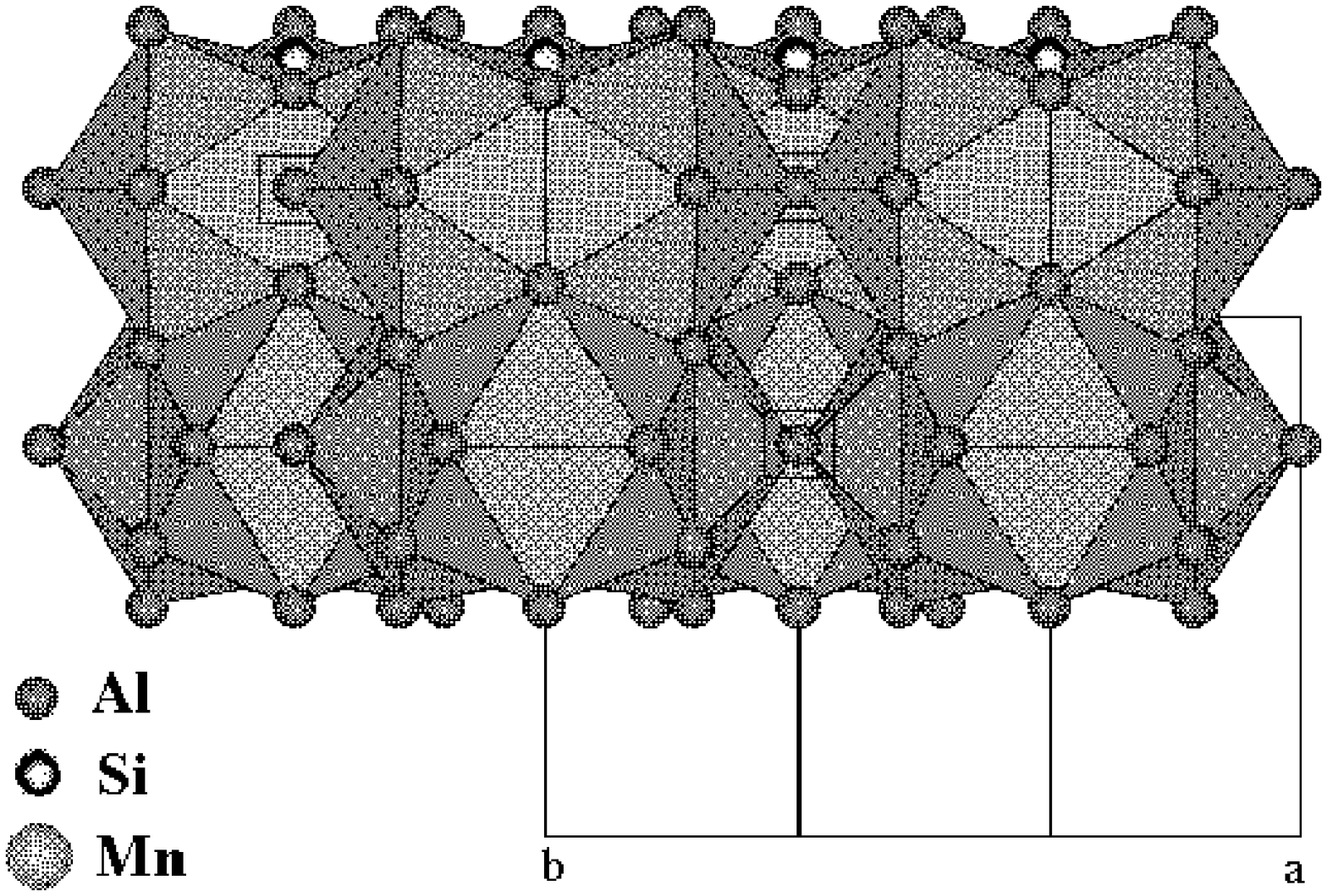,width=6cm}

\vspace{2cm}
\hspace{10.5cm}
\psfig{file=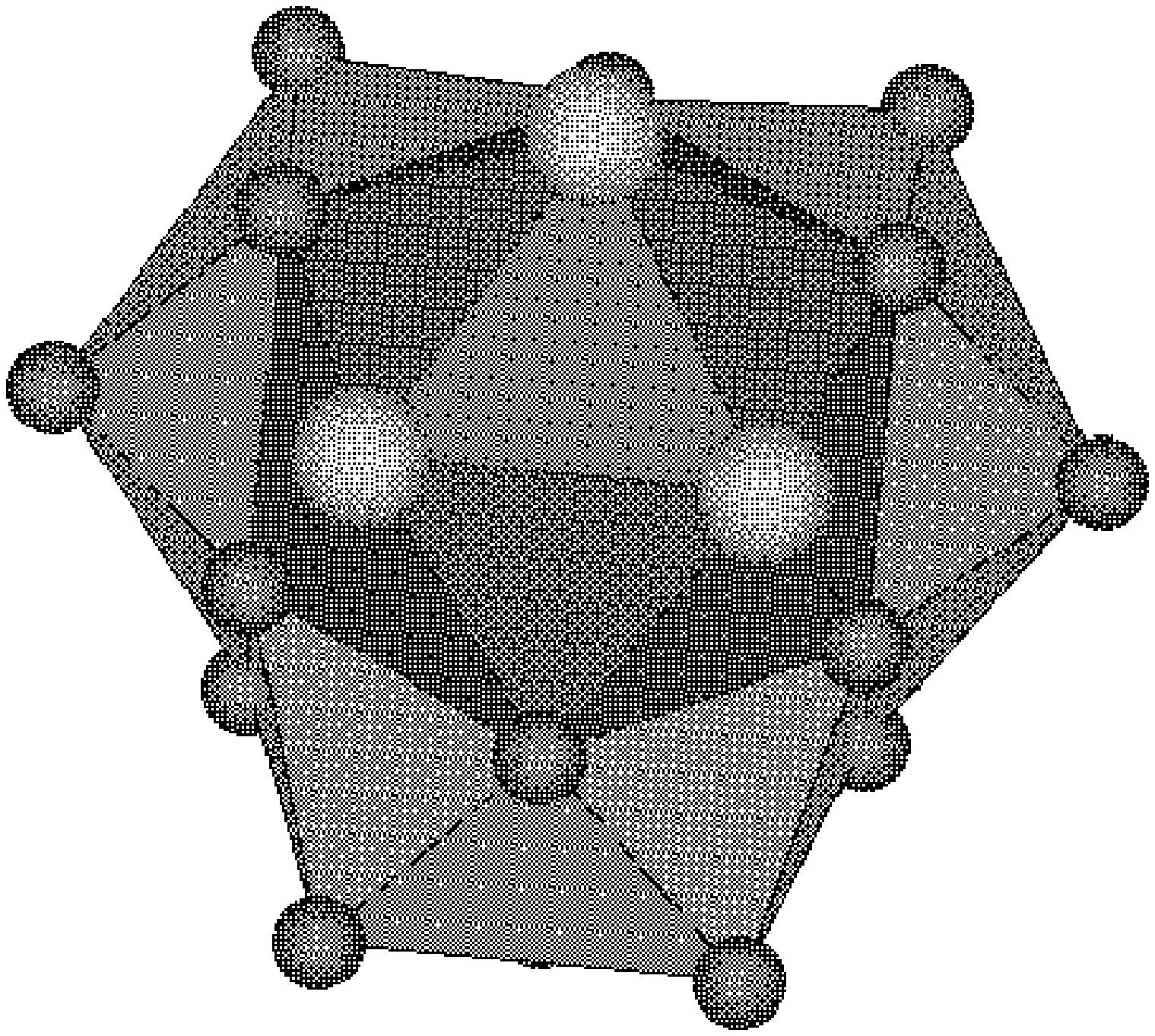,width=3cm}

\vspace{1cm}
\caption{Crystal structure of $\rm \beta\,Al_9Mn_3Si$
phase
described in terms of icosahedral clusters centered on the atoms. 
Si atoms are on  Wyckoff site (2a). 
Squares show the sites 
of the vacancy Va (site (2d)).
Top and side views of both layers at 
$z=\frac{1}{4}$, $\frac{3}{4}$ are shown.
The icosahedral environment of each Si atom
is also shown on the right low part of the figure.
}
\label{FigDessinBeta}
\end{figure}

\subsection{General aspects}
\label{SecStructure_General}

The unit cell dimensions of  $\rm \beta\,Al_9Mn_3Si$
\cite{Robinson52} and  $\rm \varphi\,Al_{10}Mn_3$
\cite{Taylor59} are  similar, with a same space group
$\rm P6_3/mmc$. 
Atomic
environment and interatomic distances
are  gathered in tables
\ref{Tab_Structure} and \ref{Tab_Distances}.

First atom neighbours of the Mn site correspond 
to 10 Al/Si + 2 Mn atoms situated at vertices of a 
distorted 
icosahedron. Such an icosahedron is considered in the 
structural representation of $\rm \beta\,Al_9Mn_3Si$ 
(figure \ref{FigDessinBeta}).
The small Al(0)-TM, Al(1)-TM and Al(2)-TM distances 
suggest a strong effect of
the sp-d hybridisation.
Each Mn has two Mn first neighbours
in Mn-triplet.
Between Mn-triplets, Mn-Mn distances are about  
$\rm 4.17\,\AA$.
Similar Mn-triplets exist also in
$\rm \mu\,Al_{4.12}Mn$
(hexagonal, $\rm P6_3/mmc$,
$\sim$\,563 atoms\,/\,unit cell) \cite{Shoemaker89}. 
$\rm \beta\,Al_9Mn_3Si$ and $\rm \varphi\,Al_{10}Mn_3$
have significant relations with the complex structures
$\rm \mu\,Al_{4.12}Mn$
and
$\rm \lambda\,Al_{4}Mn$
(hexagonal, $\rm P6_3/m$,
$\sim$\,568 atoms\,/\,unit cell) \cite{Kreiner97}
that are related to quasicrystals.
For instance, in figure 1 of Ref. \cite{Shoemaker89},
the outline of the repeated unit of $\varphi$ phase
on several parts of $\mu$ structure
is shown.
Kreiner and Franzen \cite{Kreiner95,Kreiner97}
showed that the I3-cluster, a structure unit of
three vertex connected icosahedra, is the
basic building block of a large number of
intermetallic phases related to $i$-Al-Mn-Si such as 
$\rm \alpha \,Al_9Mn_2Si$,
$\rm \mu\,Al_{4.12}Mn$ and $\rm \lambda\,Al_{4}Mn$.
Note that the environment of Mn are also close to those found in
$\rm \alpha$\,Al-Mn-Si approximant
\cite{Cooper66,Guyot85,ElserH85}.

\subsection{Vacancies}
\label{SecStructure_Vancancy}

The hexagonal structure of $\rm Al_5Co_2$ \cite{Newkirk61}
is almost 
isomorphic with $\beta$ and $\varphi$ where Co
replace Mn and Va sites (table \ref{Tab_Structure}).
These phases  have  similar atomic
sites and first-neighbours distances
(table~\ref{Tab_Distances}).
However, a major difference is that 
the site (2d) is empty (Va) in $\beta$ and $\varphi$
whereas it is
occupied by cobalt (Co(0)) in $\rm Al_5Co_2$.
It is thus interesting to understand
why this vacancy is maintained in
$\beta$ and $\varphi$ crystals?
As first-neighbour distances around Mn
in  $\varphi$ and
$\beta$ are similar to those around Co
in $\rm Al_5Co_2$, and that
Va-Al distances in $\beta$ and $\varphi$
are very close of Co(0)-Al distances
a vacancy can not be explained from steric encumbering.
The environment of Va forms a
Tri-capped trigonal prism
(3 Al(1) and 6 Al(2)).

The same environment is also found in
$\rm \mu\,Al_{4.12}Mn$ \cite{Shoemaker89} and
$\rm \lambda\,Al_{4}Mn$ \cite{Kreiner97}.
But in $\mu$ and $\lambda$, this site is occupied
by a Mn atom (Mn(1) in (2b) in $\rm \mu\,Al_{4.12}Mn$
and Mn(1) in (2d) in $\rm \lambda\,Al_{4}Mn$).
In $\mu$ and $\lambda$, 
the first-neighbour distances Mn(1)-Al are
2.35$-$2.48\,$\rm \AA$, which are similar 
to Va-Al first-neighbour distances 
in $\beta$ and $\varphi$.

In the following, it is shown that the presence (or not) 
of such a vacancy in $\beta$ and $\varphi$ can
be explained on account of the medium 
range atomic order because of  strong Mn-Mn pair 
interaction up to medium range distances 
(more than 5\,$\rm \AA$).

\subsection{Role and position of Si atoms}

The role of Si in Al based quasicrystals
and related phases
is known to have an important effect.
Unstable quasicrystals are obtained in Al-Mn system,
whereas stable quasicrystals are formed when a small proportion
of Si is added \cite{Berger94}.
Similar stabilising effects occurs for
$i$-Al-Cu-Cr-Si \cite{Khare01} and for approximants
$\alpha$\,Al-Mn-Si \cite{Berger94},
1/1\,Al-Cu-Fe-Si \cite{Quivy96,Takeuchi00},
$\alpha$\,Al-Re-Si \cite{Tamura01}.
The number of valence electrons are 3 (4) per Al (Si) atom.
With respect to a  Hume-Rothery condition for
alloying ($2k_F\simeq K_p$), it is 
possible that a substitution
of a small quantity of Si increases the $e/a$ ratio
in better agrement with $2k_F\simeq K_p$.

Experimentally
Al and Si atoms have not been distinguished in 
$\rm \beta\,Al_9Mn_3Si$.
However, Robinson \cite{Robinson52} has proposed to 
consider Si 
in  (2a) because
the interatomic distances between an atom on
sites (2a) and its six neighbouring Al atoms is less
than between any other pairs of Al atoms in $\beta$ structure
(table\,\ref{Tab_Distances}).
But, from a comparison of $\beta$ and $\varphi$,
Taylor\,\cite{Taylor59} has suggested 
that Si atoms should be preferentially in
(12k) with Al atoms instead of (2a).

In section \ref{SecStability}, we give arguments
form {\it ab initio} calculations to
understand the effect of Si on the Hume-Rothery stabilisation
and to conclude that  Si are likely in (2a).

%
%
\section{First-principles calculations of the electronic
structure}
\label{SecLMTO}

\subsection{LMTO procedure, treatment of Si}
\label{SecLMTOprocedure}

Electronic structure determinations were 
performed in the frame-work of the local
spin-density approximation (LSDA) \cite{Barth72} by using the
{\it ab initio} Linear Muffin Tin Orbital method (LMTO) in an Atomic
Sphere Approximation (ASA) \cite{Andersen75,Skriver84}.
The space is divided into
atomic spheres and
interstitial region 
where the potential is spherically symmetric
and flat, respectively. 
Sphere radii were chosen so that the total volume of spheres
equals that of the solid.
For vacancies (Va)  empty spheres were introduced
in (2d).
The sphere radii are
$R_{Si/Al(0)}=1.37$\,$\rm \AA$,
$R_{Al(1)}=R_{Al(2)}=1.53$\,$\rm \AA$,
$R_{Mn}=1.34$\,$\rm \AA$,
$R_{Va}=1.04$\,$\rm \AA$ for $\beta$ phase,
and
$R_{Al(0)}=1.38$\,$\rm \AA$,
$R_{Al(1)}=R_{Al(2)}=1.55$\,$\rm \AA$,
$R_{Mn}=1.35$\,$\rm \AA$,
$R_{Va}=1.05$\,$\rm \AA$ for $\varphi$ phase.
As these structures are metallic and rather
compacts, it was
found that a small change of the sphere radii
does not modify significantly the results.

Neglecting the spin-orbit coupling,
a scalar relativistic LMTO-ASA code,
was used
with combined corrections for ASA
\cite{Andersen75,Skriver84}.
The {\bf k}
integration in a reduced Brillouin zone was performed 
according to
the tetrahedron method
\cite{Jepsen71} 
in order to calculate the electronic density of states (DOS).
The final step of the self-consistent procedure and
the DOS calculation were performed with 4416 {\bf k} points in the
reduced Brillouin zone. 
With an energy mesh equals to $\Delta E=0.09$\,eV,
calculated DOSs
do not exhibit significant differences
when the number of {\bf k} points increases from 2160 to 4416.
Thus,
the structure in the DOS larger than 0.09\,eV are
not artefacts in  calculations.
Except in section \ref{SecMagnetism}, 
the LMTO DOSs calculations were performed 
without polarised spin (paramagnetic state).

The LMTO-ASA basis includes all angular moments up
to $l=2$ and the valence states are Al (3s, 3p, 3d), Mn (4s, 4p,
3d), Co (4s, 4p, 3d), Si (3s, 3p, 3d) and Va (1s, 2p, 3d)
\footnote{In ASA approximation, orbitals are introduced 
in vacancies in order to yield a good expansion 
of the LMTO orbitals out of  atomic spheres.}.
In order to analyse the position of Si atoms in the $\beta$ phase,
we performed
calculations for  $\rm \beta (Al,Si)_{10}Mn_3$ where
the Si atoms occupied randomly the
Al sites. 
In this case an average atom
named (Al,Si) was considered (virtual crystal approximation).
In the LMTO-ASA procedure this atom is simulated 
with nuclear charge $Z=(1-c)Z_{Al} + cZ_{Si}$, where $c$ is the
proportion of Si atoms, and $Z_{Al}=13$, $Z_{Si}=14$ are the nuclear
charge of Al, Si, respectively.
Such a calculation can be justified as the main difference between Al
and Si is the number of valence electrons.
It was checked that the LMTO-ASA total energy of pure Al and pure Si are
almost equal to this calculated with the average (Al,Si) atom
with $c=0.01$ and $c=0.99$, respectively.
Three possibilities were considered for the $\beta$ phase:
(i) the phase named $\rm \beta\,Al_9Mn_3Si$ where
Si are on site (2a) and Al, on
site (6h) and site (12k);
(ii) the phase $\rm \beta\;I$-$\rm (Al,Si)_{10}Mn_3$ where
Si atoms substitute for some Al (on sites
(2a), (6h) and (12k));
(iii) the phase $\rm \beta\;II$-$\rm (Al,Si)_{10}Mn_3$ where
Si atoms substitute for some Al(2) (site (12k)).
Same sphere radii for these three cases were input.

\subsection{General aspects of the density of states (DOS)}

\label{SecStability}

\begin{figure}
\begin{center}
\psfig{file=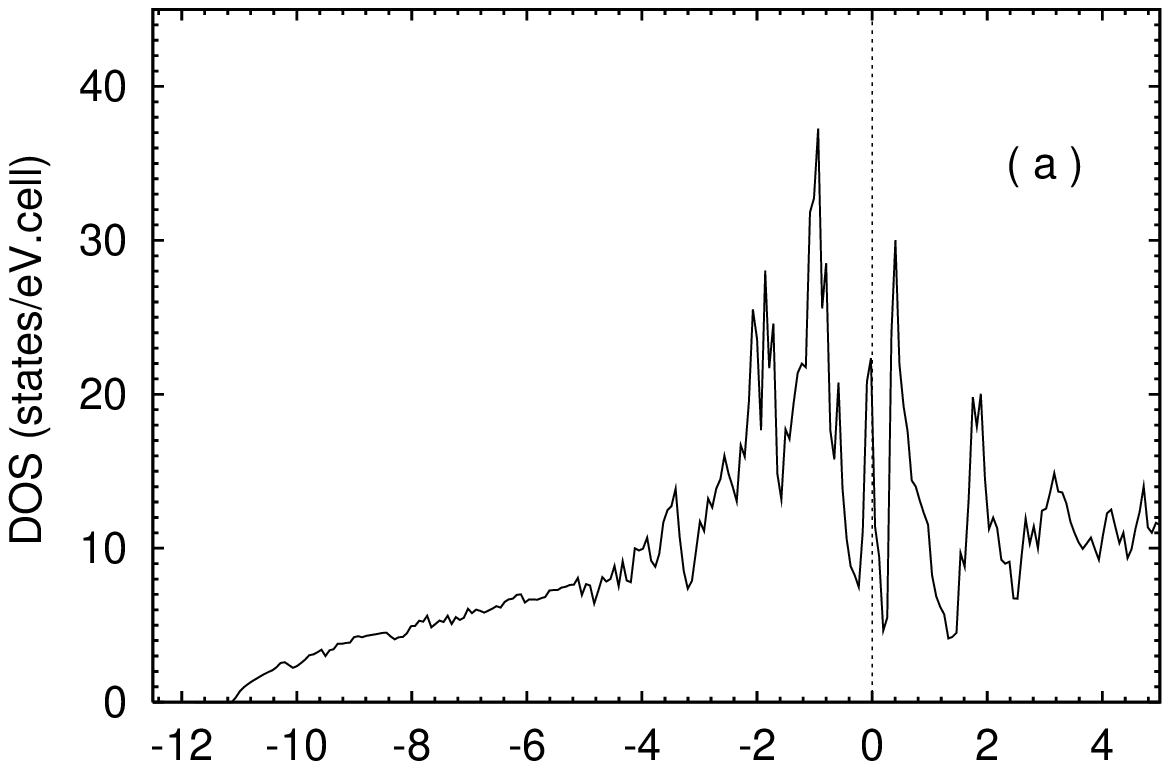,width=10cm}

\vspace{-6.5cm}\hspace{-4cm}
\psfig{file=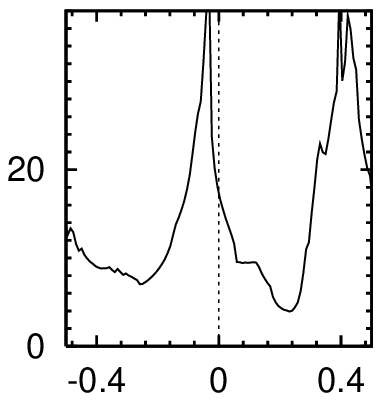,width=4.2cm}

\vspace{1.7cm}
\psfig{file=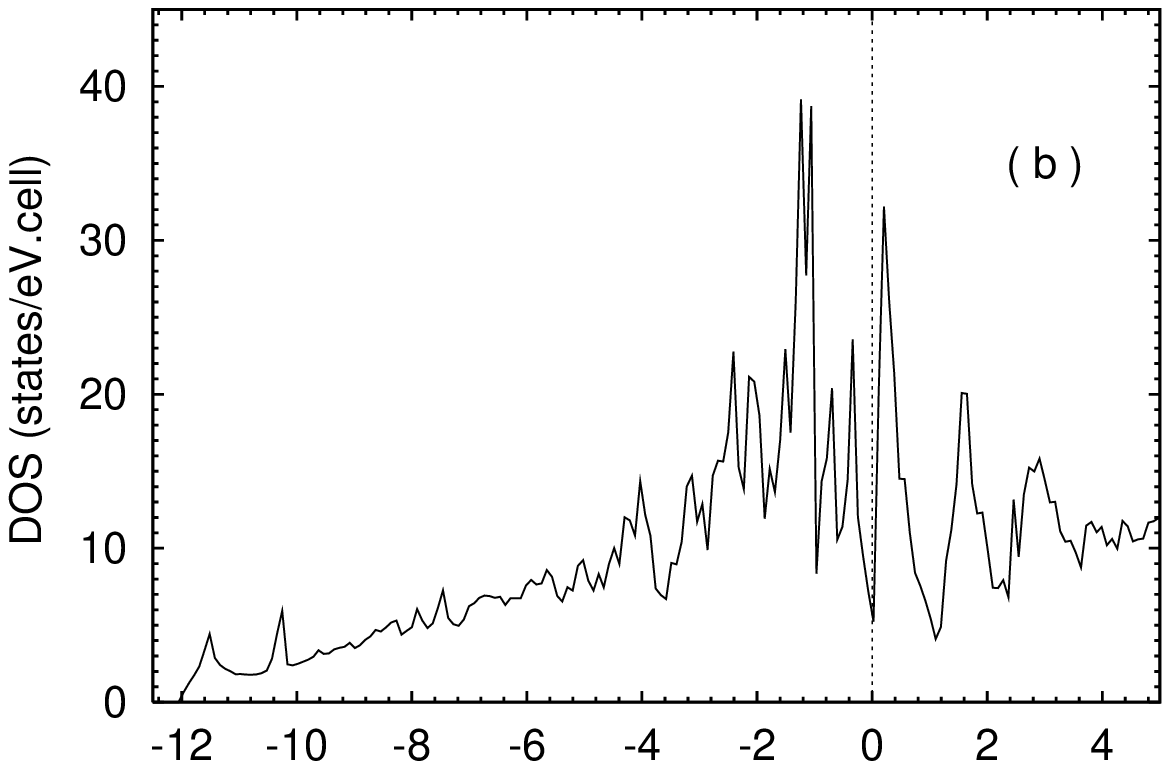,width=10cm}

\vspace{-6.5cm}\hspace{-4cm}
\psfig{file=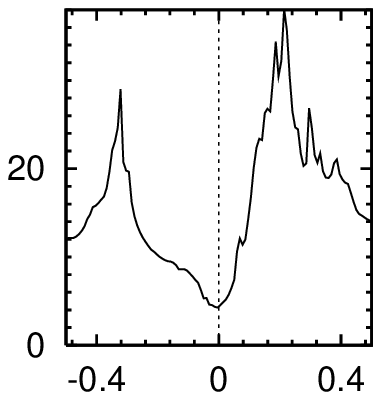,width=4.2cm}

\vspace{1.7cm}
\psfig{file=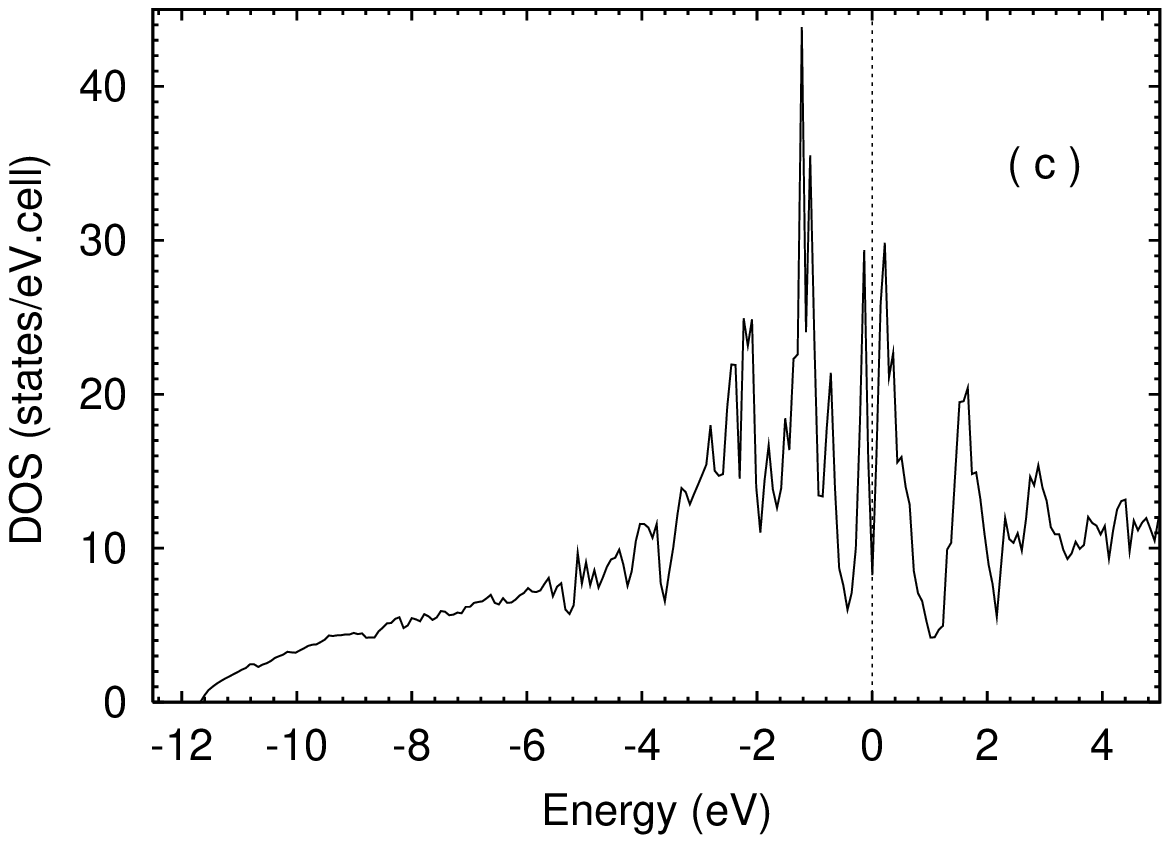,width=10cm}

\vspace{-6.5cm}\hspace{-4cm}
\psfig{file=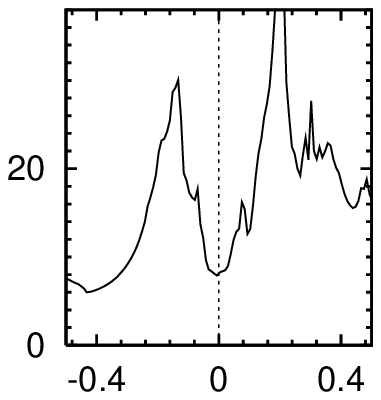,width=4.2cm}

\vspace{2.5cm}
\caption{Total density of states (DOS) calculated by
LMTO-ASA method in (a)
$\rm \varphi\,Al_{10}Mn_3$,
(b) $\rm \beta\,Al_9Mn_3Si$, and
(c) $\rm \beta\,I$-$\rm (Al,Si)_{10}Mn_3$.
Details of the total DOSs around $E_F$ are given in inserts.
$E_F=0$.
The DOS in $\rm \beta\,II$-$\rm (Al,Si)_{10}Mn_3$
is almost the same as that in
$\rm \beta\,I$-$\rm (Al,Si)_{10}Mn_3$.}
\label{FigDOStotale}
\end{center}
\end{figure}

Total energy self-consistent calculations were performed
for different volumes, with isotropic volume changes i.e.
the ratio $c/a$ is constant and equal to the experimental value
(table~\ref{Tab_Structure}). 
The atomic positions were not relaxed. 
Minima of  energy were obtained for a lattice parameter
$a$ equal to 
7.41\,$\rm \AA$ for  $\rm \beta\,Al_9Mn_3Si$, 
7.41\,$\rm \AA$ for $\rm \beta\,I$-$\rm (Al,Si)_{10}Mn_3$,
7.42\,$\rm \AA$ for $\rm \beta\,II$-$\rm (Al,Si)_{10}Mn_3$,
and 7.43\,$\rm \AA$ for $\rm \varphi\,Al_{10}Mn_3$.
These values correspond 
within $1.5\,\%$ to experimental values.
Similar results have also been found in 
LMTO-ASA calculations for Al-TM alloys
with small concentration of TM elements \cite{GuyPRB95}.

\begin{figure}
\begin{center}

\psfig{file=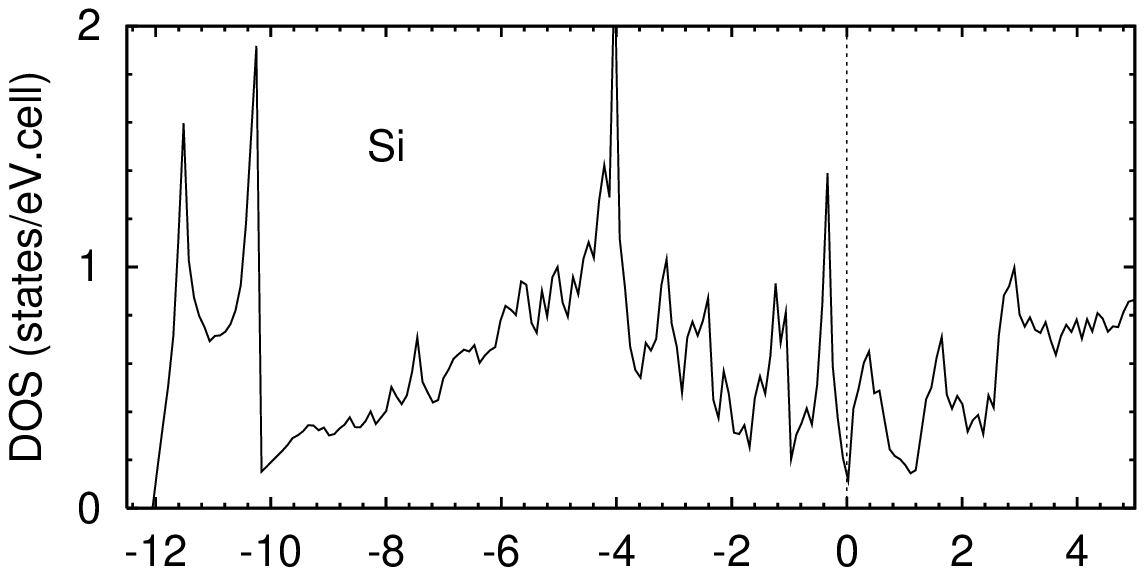,width=7cm}
\hspace{-0.85cm}\psfig{file=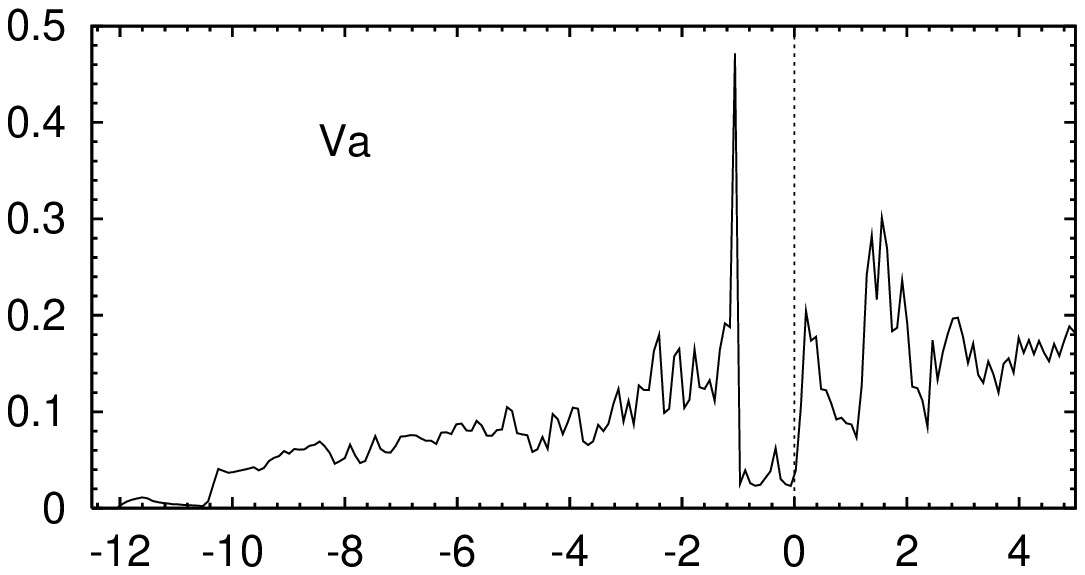,width=7.1cm}

\vspace{-0.3cm}
\psfig{file=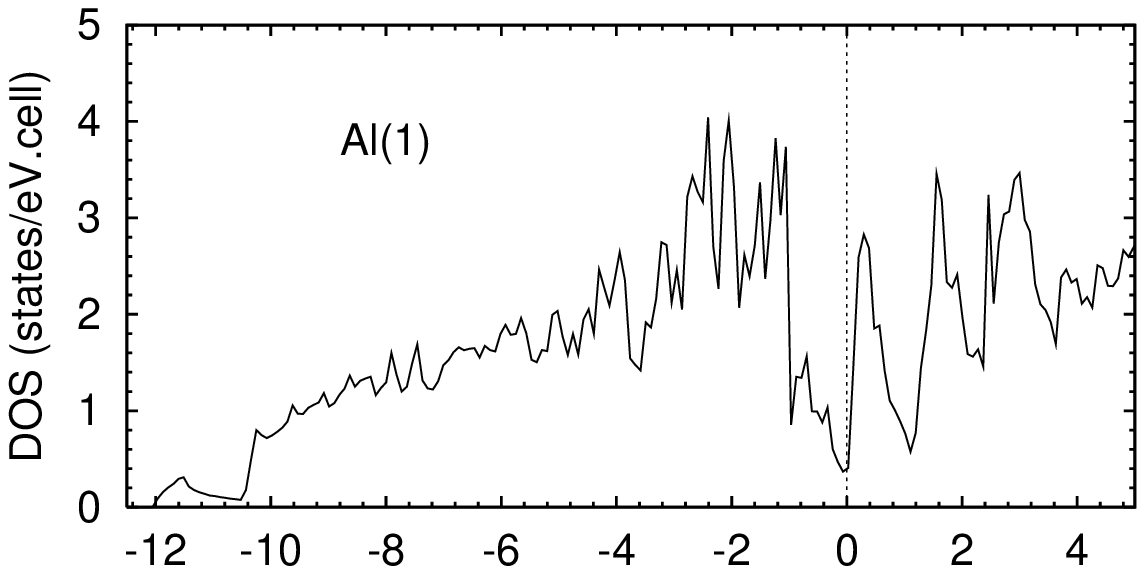,width=7cm}
\hspace{-0.7cm}\psfig{file=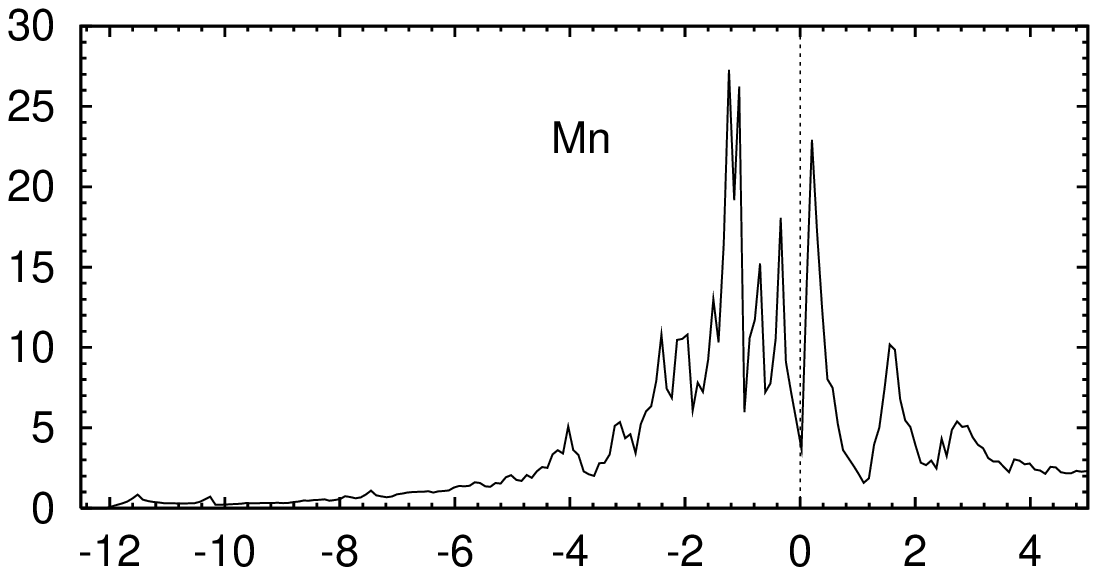,width=7cm}

\vspace{-0.3cm}
\psfig{file=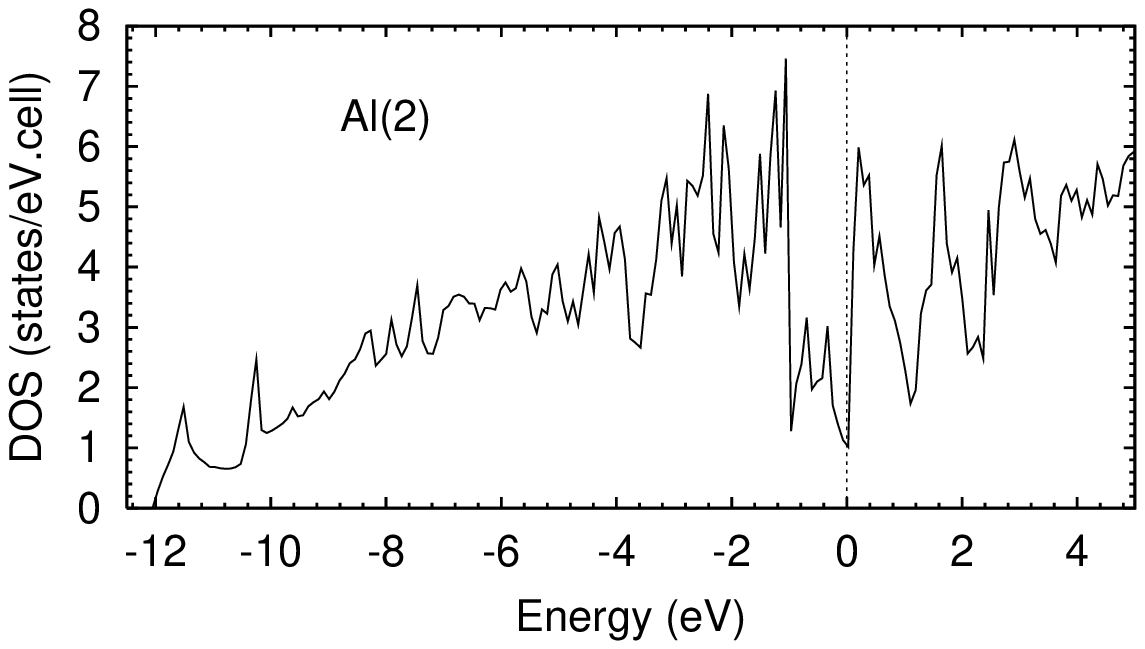,width=7cm}
\hspace{-0.7cm}\psfig{file=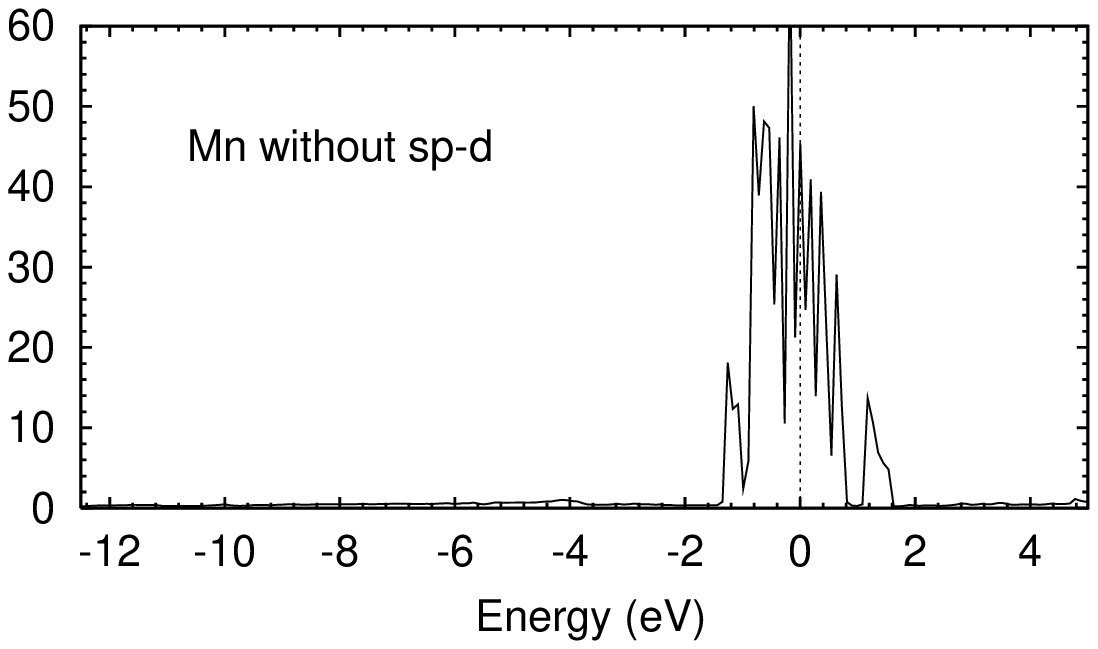,width=7cm}

\vspace{-6.5cm} \hspace{3.5cm}
\psfig{file=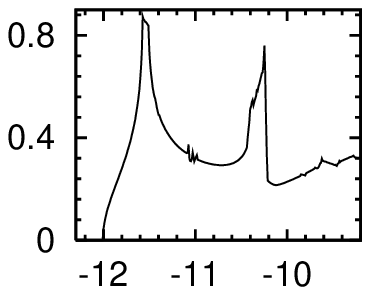,width=3cm}

\vspace{4.1cm}
\caption{Local DOS performed by
LMTO-ASA method in $\rm \beta\,Al_9Mn_3Si$
phase. 
Si are in (2a). 
The local Mn DOS calculated without sp-d
hybridisation is also drawn (see text). $E_F=0$.}
\label{FigDOSlocalBeta}
\end{center}
\end{figure}

The total DOSs
in $\beta$ and $\varphi$ phases
(figure~\ref{FigDOStotale}), are very similar.
Local DOSs in $\beta$ are also shown in
figure~\ref{FigDOSlocalBeta}.
Except for low energies (less than $-10\,$eV), the total DOS
in $\beta$ does not depend on the Si position.
The parabola due to the
Al nearly-free states is clearly seen.
The large d
band from $-2$ up to $2$\,eV
is due
to a strong sp-d hybridisation 
in agreement with
experimental results 
\cite{Belin97,Mori91,Belin92,Dankhazi93}
and with first-principles
calculations
on  Al-TM crystals and quasicrystals
\cite{GuyPRB95,FujiAlMnSi,Hafner98a}.

\begin{figure}
\begin{center}
\psfig{file=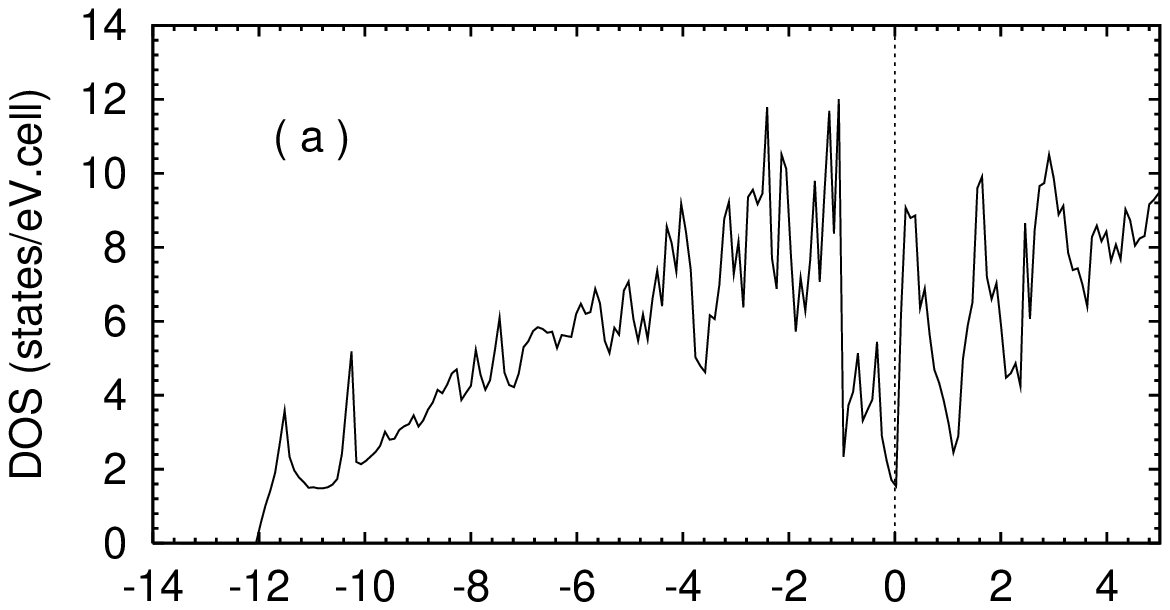,width=7cm}
\hspace{-0.85cm}
\psfig{file=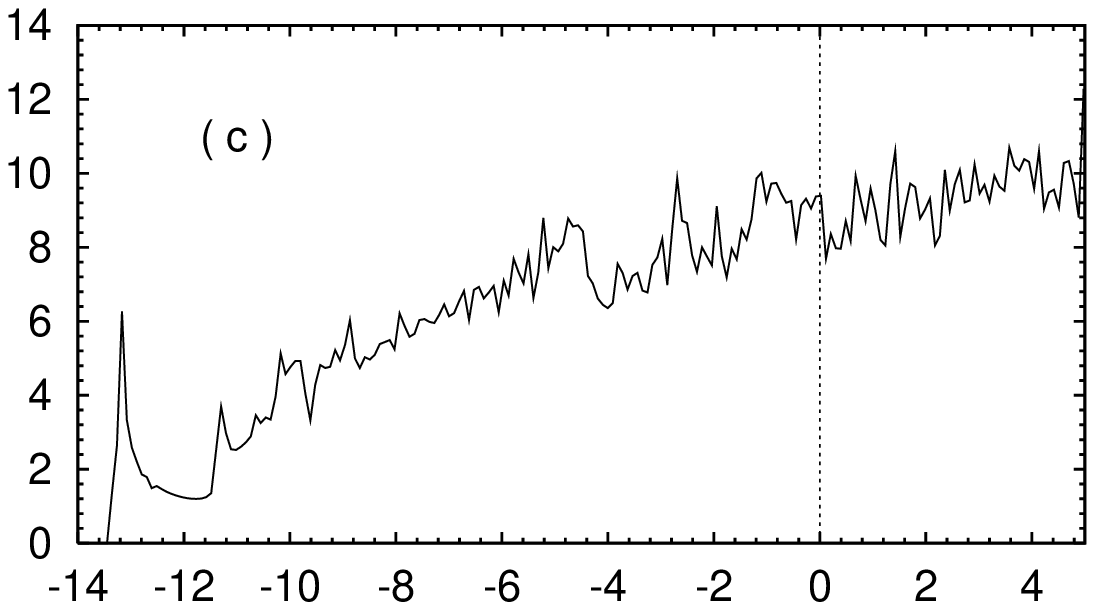,width=7cm}

\vspace{-0.5cm}
\psfig{file=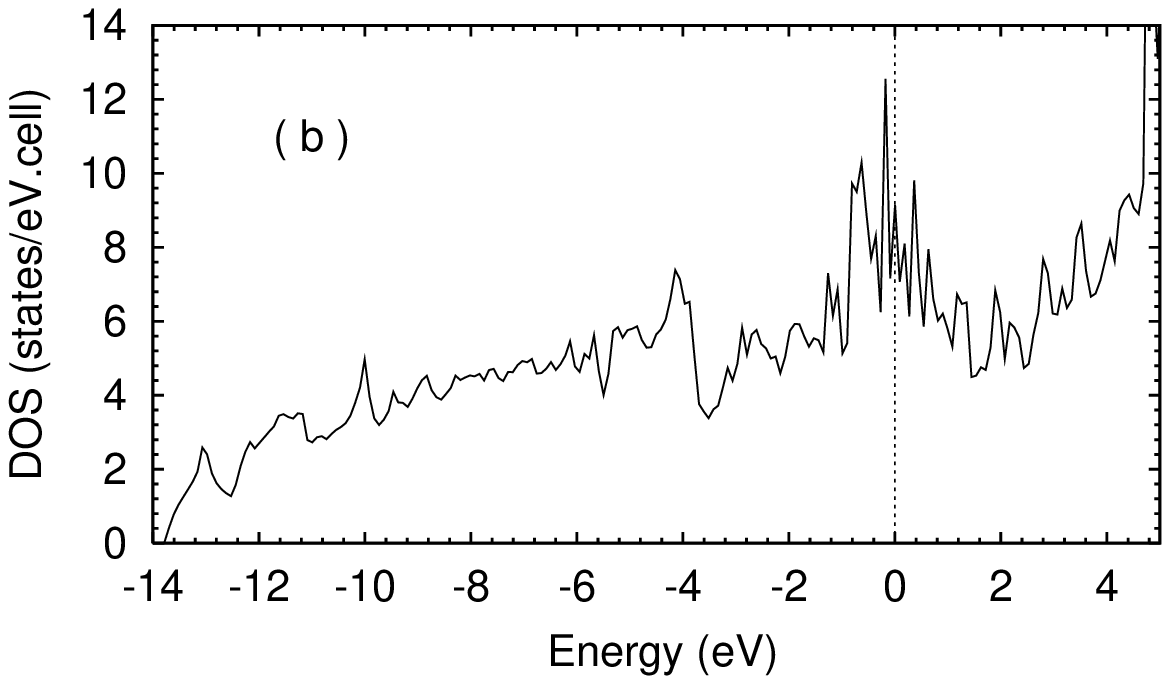,width=7cm}
\hspace{-0.85cm}
\psfig{file=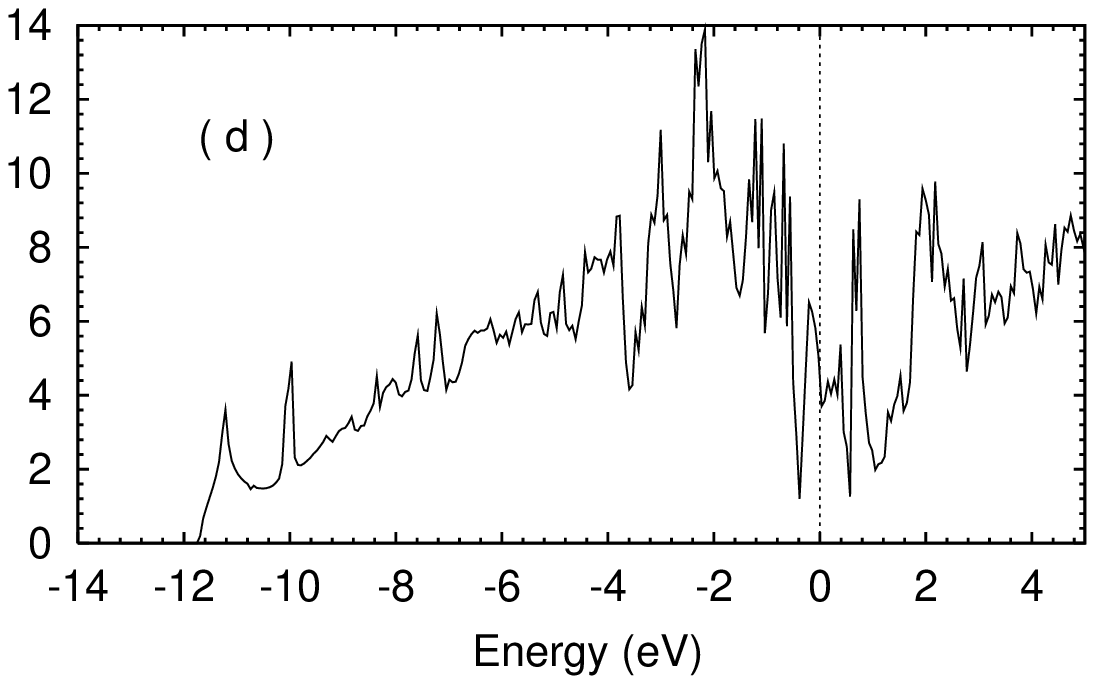,width=7cm}

\caption{local \{Al + Si\} DOS performed
by LMTO-ASA method in $\rm \beta\,Al_9Mn_3Si$
phase: (a) calculated including sp-d hybridisation,
(b) calculated without sp-d hybridisation.
(c) local \{Al + Si\} DOS in hypothetical
$\rm \beta\,Al_9Al_3Si$ and (d) in  hypothetical
$\rm \beta\,Al_9Mn_4Si$. $E_F=0$.}
\label{FigDOSAlSi}
\end{center}
\end{figure}

The sum of local DOSs on Al and Si atoms, shown
in figure~\ref{FigDOSAlSi}(a),
is mainly sp DOS.
As expected for a Hume-Rothery stabilisation,
it exhibits a wide pseudogap near $E_F$ due to
electron scattering  by Bragg planes of
a predominant pseudo-Brillouin zone 
(i.e. the pseudo-Brillouin zone close to 
the Fermi surface). 
Its width
of about 1\,eV is of the same order
of magnitude to this found
in Al-Mn icosahedral
approximants 
\cite{GuyPRB95,FujiAlMnSi,GuyPRB94_AlCuFe,Hafner98a}.
The large pseudogap in \{Al\,$+$\,Si\} DOS is
meanly characteristic of a
p-band at this energy, 
but  the pseudogap in the total DOS is narrower.
Therefore, the d states of 
Mn atoms must fill up partially the pseudogap.
Nevertheless, as it is shown
in the following that the pseudogap 
in \{Al\,$+$\,Si\} DOS results from Mn sub-lattice effect.

Spiky total DOS 
where obtained for the studied phases as it has
been found in
LMTO DOS of icosahedral small  approximants
(for instance $\rm \alpha\,$Al-Mn-Si
\cite{FujiAlMnSi},
1/1\,Al-Cu-Fe \cite{GuyPRB94_AlCuFe}, 1/1\,Al-Pd-Mn
\cite{Krajvci95}).
In fact this is a consequence of
a small electron velocity  (flat dispersion
relations) which contributes to
anomalous electronic transport properties
\cite{FujiAlMnSi,GuyPRB94_AlCuFe,MayouPRL00}.
Such properties are not specific 
of quasicrystals as they are also observed 
in many crystal related
to quasicrystals,
therefore it does not only come  from the long
range quasiperiodicity. 
They are
also associated
with local and medium range atomic order
that are related to quasiperiodicity.
Indeed, it has been shown \cite{GuyPRB97} 
that fine peaks in the DOS could come
from electron confinement in atomic
clusters characteristic \cite{Gratias00}
of the quasiperiodicity.
This is not in contradiction with a Hume-Rothery mechanism 
because this
tendency to localisation has a small effect on the
band energy \cite{GuyPRB97}.
Whether spiky DOSs exist in quasicrystals
or not is however much debated
experimentally \cite{Stadnik97,Dolinsek00,Stadnik01} 
and theoretically
(Ref \cite{Zijlstra00} and references therein) and
the present calculation does not give 
answer to this question for the case of 
quasicrystals.
But, in case of $\beta$ and $\varphi$
crystals, we
checked that 
structures in the DOS with an energy scale larger
than 0.09\,eV are not artifacts in calculation
as they do not depend on  the
non-physical parameters in the LMTO
procedure (number of {\bf k} points,
see section \ref{SecLMTOprocedure}).

\subsection{Analysis of Si effect}

From LMTO band energy calculated
with fixed atomic positions and 
composition,
the lowest band energy is obtain
when Si atoms are in (2a) in $\rm \beta\,Al_9Mn_3Si$.
The difference in band energy between $\rm \beta\,Al_9Mn_3Si$
and $\rm \beta\;I$-$\rm (Al,Si)_{10}Mn_3$ is
$+5$\,eV\,/\,unit cell. 
The same order of magnitude is obtained
between $\rm \beta\,Al_9Mn_3Si$ and
$\rm \beta\;II$-$\rm (Al,Si)_{10}Mn_3$.
These {\it ab initio} results shows that Si are
in (2a) and  not mixed with
Al.
A comparison between the band energy of
$\beta$ and $\varphi$ phases cannot be made
because their compositions are different.

In the vicinity of
$E_F$, the total DOSs in $\beta$ and $\varphi$ are
very similar
except $E_F$ positions (figure \ref{FigDOStotale}).
At $E_F$, the DOS is 
$5.6\,$states$/$eV.cell in $\beta$, 
$16.0\,$states$/$eV.cell in $\varphi$.
The small amount of Si 
increases the average valence $e/a$ in
$\beta$. 
In a rigid band like model,
this $E_F$ shifts   up to the
minimum of the pseudogap,
and in  Hume-Rothery mechanism, the band energy
is minimised.
This difference allows one to understand why 
$\rm \beta\,Al_9Mn_3Si$ 
phase is stable
whereas $\rm \varphi Al_{10}Mn_3$ phase is metastable.

Differences between
$\rm \beta\,(Al,Si)_{10}Mn_3$
where Si atoms are mixed with
Al atoms,  and $\rm \beta\,Al_9Mn_3Si$
where Si atoms are in (2a) can also be understood 
from LMTO DOSs.
Indeed, two bonding peaks are present at low
energies ($-11.5$\,eV and $-10.3\,$eV, 
figure~\ref{FigDOSlocalBeta})
in the local Si DOS of $\rm \beta\,Al_9Mn_3Si$, which is no more
nearly-free states
(each Si atom has 6 Al(2) and 6 Mn first neighbours
(table~\ref{Tab_Distances})).
The close
proximity between Si and Mn and the presence of 
a bonding peak in the partial
Mn DOS suggest that the Si-Mn bond  
is rather covalent and thus
increases the stability of $\beta$ phase when Si is on site
(2a).

\subsection{Effects of the d state of the  transition-metal (TM) atoms}
\label{Sec_spd}

In this part the origin of the pseudogap is analysed from 
LMTO calculations for
hypothetical phases derived
from $\rm \beta\,Al_9Mn_3Si$.
Three points are successively considered 
$(i)$ the strong effect of the
sp-d hybridisation on the pseudogap, 
$(ii)$ the role of the Mn position which explains
the origin of the vacancy (Va) in $\beta$ and $\varphi$, 
$(iii)$ and the
great effect of Mn-Mn medium range interaction up to $\rm 5\,\AA$.

\subsubsection*{(i) Role of the sp-d hybridisation 
on the pseudogap\\ \\}

Self-consistent
LMTO calculation where performed 
without sp-d hybridisation by setting to zero
the corresponding terms of the hamiltonian matrix 
\cite{Duc92}.
Such a calculation is physically meaningful
because the d TM states are mainly localised in the TM sphere and the
sp Al states are delocalised.
In figure \ref{FigDOSlocalBeta}, local Mn DOS (mainly d states),
is drawn for two cases:
with sp-d hybridisation and without sp-d hybridisation.
The comparison between these local 
DOSs shows that the sp-d hybridisation
increases the width of d band.
This confirms a strong sp-d hybridisation.
The local \{Al + Si\} DOS (mainly sp DOS) is also strongly
affected by a sp-d hybridisation.
As a matter of fact the pseudogap disappears in the calculation
without sp-d hybridisation (figure~\ref{FigDOSAlSi}(b)).

For Hume-Rothery alloys containing TM elements,
a stabilisation
mechanism is more complex than in sp alloys because
of a strong sp-d
hybridisation in the vicinity of $E_F$.
Al and Si atoms, which have a weak potential, scatter sp electrons
by a potential $V_B$ almost energy independent.
This leads to the so-called diffraction of electrons
by Bragg planes in sp alloys.
But the potential of Mn atoms
depends on the energy. It is strong for 
energies around $E_d$, and creates a d resonance of the
wave function that scatters also  sp states.
The effect is analysed in detail from a model hamiltonian 
in section \ref{BraggPotential}.
LMTO calculation was performed on
hypothetical $\rm \beta\,Al_9Al_3Si$ constructed by
putting Al in place of Mn in $\rm \beta\,Al_9Mn_3Si$
in order to confirm (or not) the previous analysis. 
The resulting 
total DOS (mainly sp DOS) of the hypothetical phase has
no pronounced pseudogap (figure \ref{FigDOSAlSi}(c)).
But
there are many small depletions that might come from
diffractions by Bragg planes. It shows that classical
diffractions by Bragg planes by a weak potential $V_B$
can not explain a pseudogap close to $E_F$ in
$\rm \beta\,Al_9Al_3Si$ and
$\rm \varphi\,Al_{10}Mn_3$.

\subsubsection*{(ii) Effect of the Mn position, 
origin of the vacancy\\ \\}
\label{LMTO_Vacancy}

\begin{figure}
\begin{center}
\hspace{2cm}$\rm Al_5Co_2$  \hspace{3.cm}
hypothetical $\rm \beta\,Al_9Mn_4Si$

\vspace{0.3cm}
\psfig{file=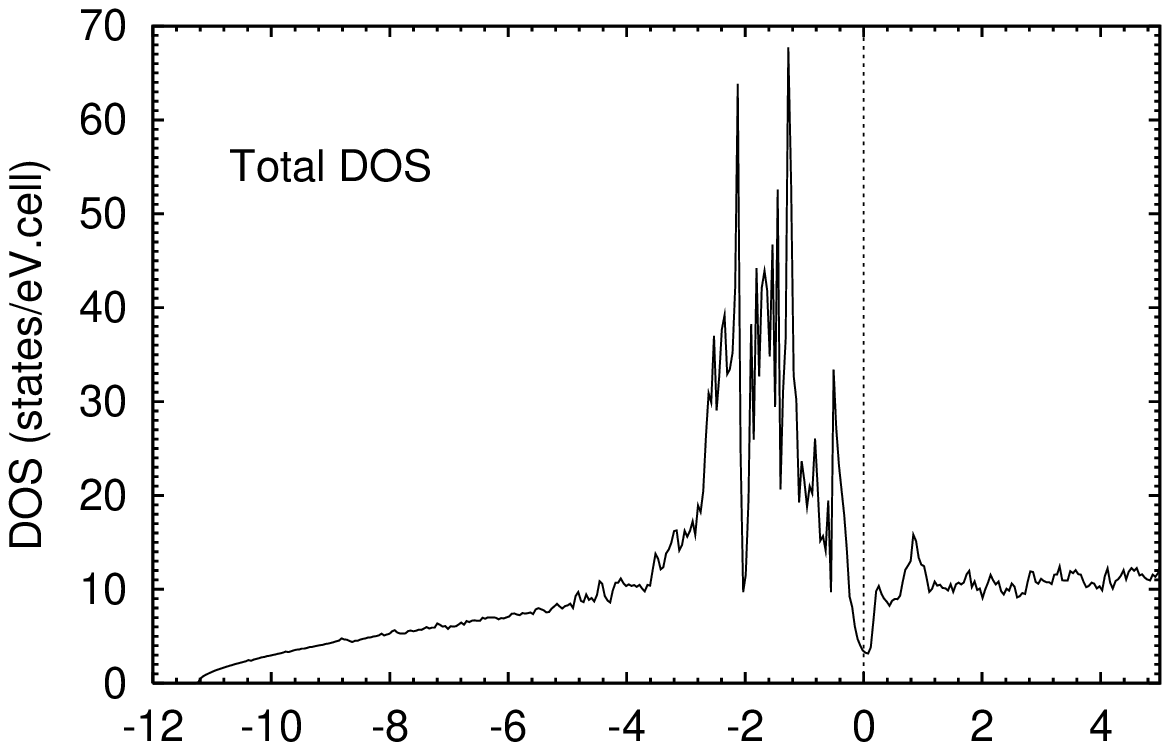,width=7cm}
\hspace{-0.9cm}\psfig{file=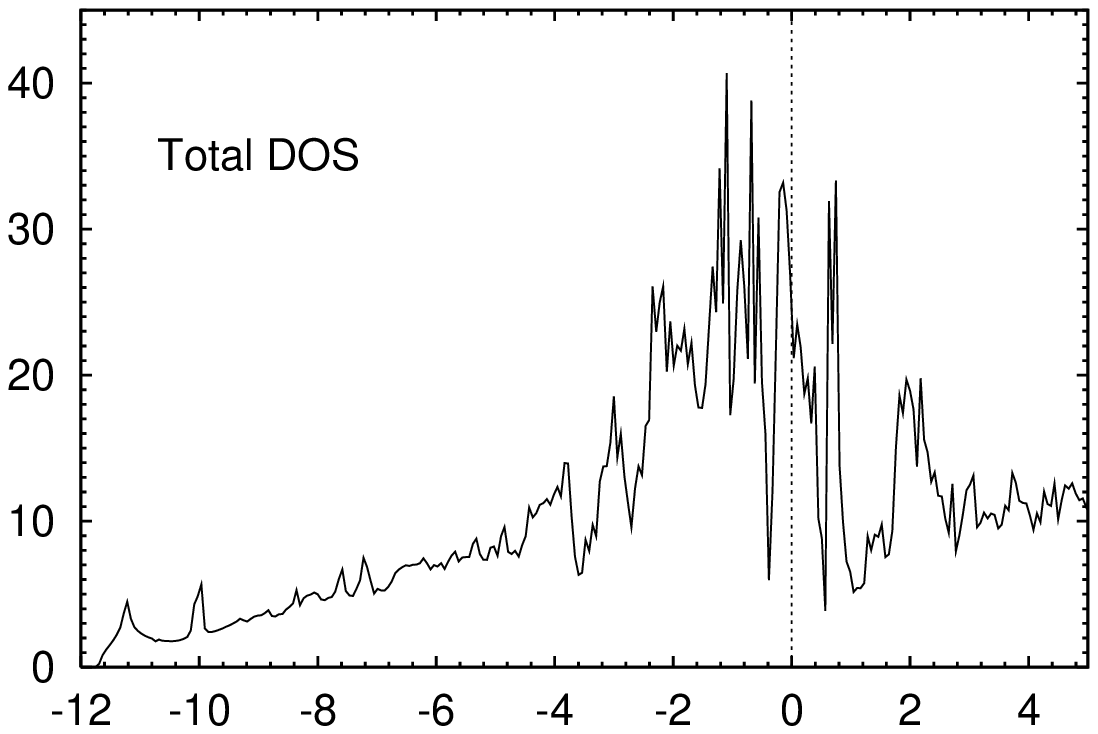,width=7cm}

\vspace{-0.2cm}
\psfig{file=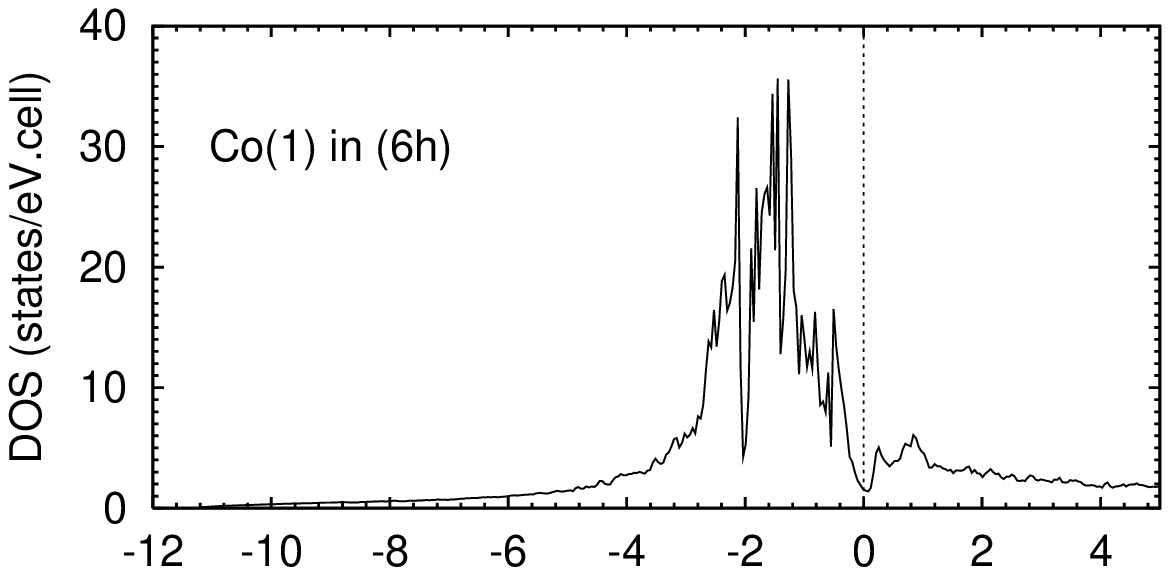,width=7cm}
\hspace{-0.9cm}\psfig{file=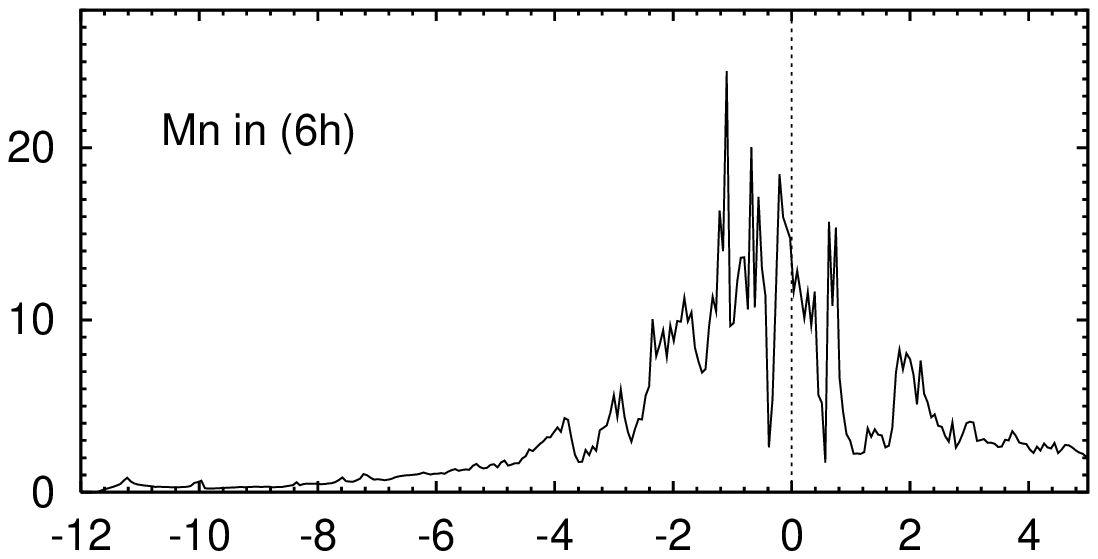,width=7cm}

\vspace{-0.2cm}
\psfig{file=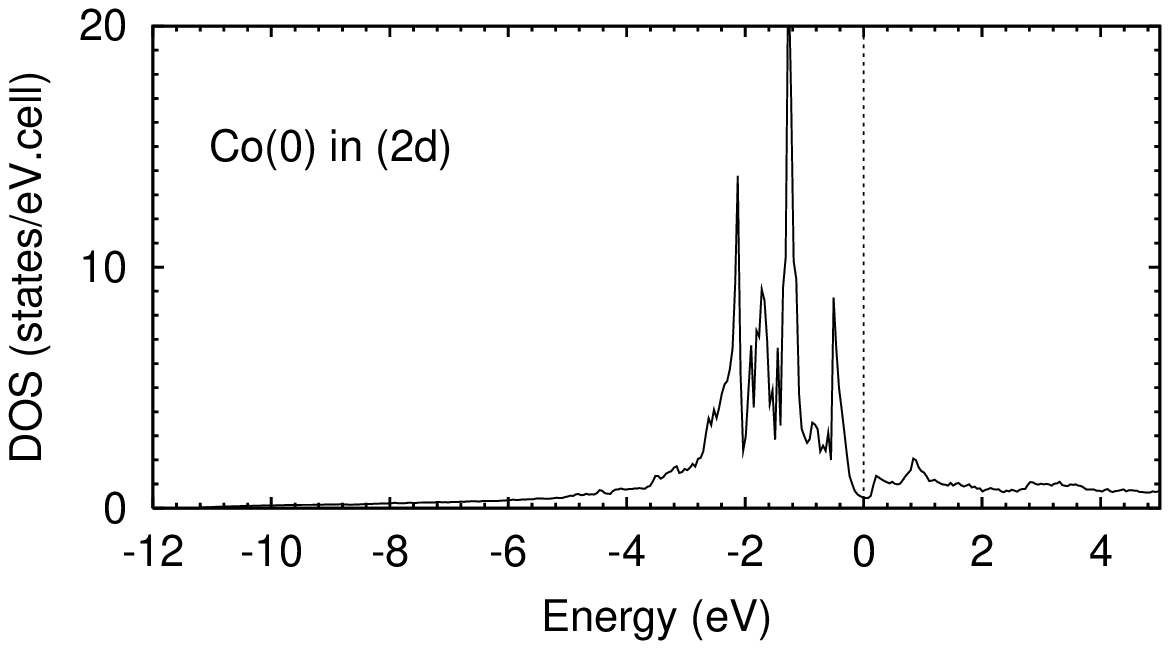,width=7cm}
\hspace{-0.9cm}\psfig{file=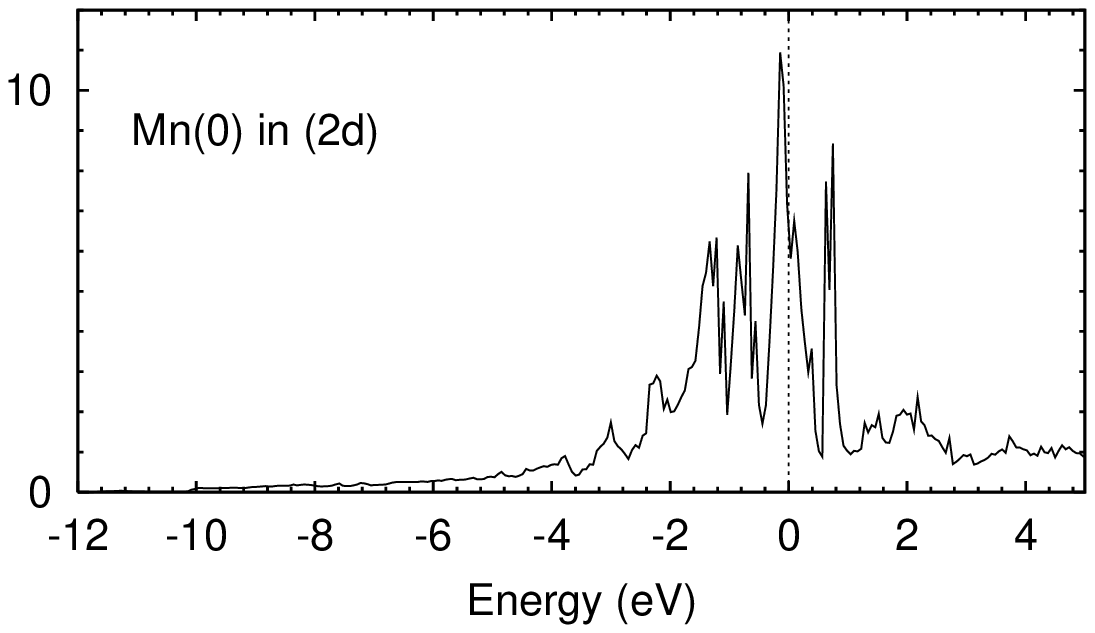,width=7cm}

\caption{Total DOS and local TM DOSs performed
by LMTO-ASA method in
$\rm Al_5Co_2$  and hypothetical
$\rm \beta\,Al_9Mn_4Si$. The $\rm \beta\,Al_9Mn_4Si$
phase is built from $\rm \beta\,Al_9Mn_3Si$
(table \ref{Tab_Structure}) 
by replacing the vacancy in (2d) 
by  Mn atom (Mn(0)). 
Other Mn are in (6h). $E_F=0$.}
\label{FigDOSAl5Co2}
\end{center}
\end{figure}

As explained in section \ref{SecStructure}, 
a particularity of $\beta$ and
$\varphi$ structures is a vacancy in (2d).
This is the main difference with the $\rm Al_5Co_2$ structure
(table~\ref{Tab_Structure}).
The origin of a vacancy can not be explained from too short
near-neighbour distances (section \ref{SecStructure_Vancancy}).
Therefore, a
LMTO calculation was performed including a new Mn atom, named
Mn(0), on site (2d) in $\rm \beta\,Al_9Mn_3Si$ phase
using atomic sites and lattice parameters of
$\rm \beta\,Al_9Mn_3Si$ (table \ref{Tab_Structure}).
This hypothetical phase is named $\rm \beta\,Al_9Mn_4Si$
and
its total and  \{Al\,$+$\,Si\} DOSs  are shown 
in figures
\ref{FigDOSAl5Co2} and 
\ref{FigDOSAlSi}(d),
respectively.
The absence of pseudogap in the total DOS results of a
the great effect of Mn(0).
In \{Al\,$+$\,Si\} DOS
the pseudogap created by the scattering
of sp electrons by the sub-lattice of Mn in (6h) 
is still present.
But a large peak at $E_F$ fills up partially the pseudogap.
Consequently, $E_F$ is located in a peak due to sp(Al)-d(Mn(0))
hybridisation.
This is in fact a good example where 
a sp-d hybridisation does not induce
a pseudogap.
A similar result was obtained with an hypothetical
$\rm \varphi\,Al_{10}Mn_4$, built by putting a Mn atom in place of
the vacancy in (2d).

Total and local DOSs of
hypothetical $\rm \beta\,Al_9Mn_4Si$ and 
$\rm Al_5Co_2$ \cite{GuyPRB95,Belin97}
are compared in figure \ref{FigDOSAl5Co2}.
In spite of the near isomorphism between these structures,
their DOSs are very different. 
As there is pseudogap in $\rm Al_5Co_2$ and not in 
$\rm \beta\,Al_9Mn_4Si$, it indicates that 
Co(0) and Mn(0) act differently,
thus justifying the existence of a vacancy 
in both $\rm \beta\,Al_9Mn_3Si$
and $\rm \varphi\,Al_{10}Mn_3$ phases and not in
$\rm Al_5Co_2$.

Therefore,
the similar Wyckoff sites lead to both  anti-bonding 
or bonding peaks 
depending on the nature of the atom on the Wyckoff site (2d),
either Mn(0) or Co(0) respectively.
Since there is a great effect of  the nature 
of the TM element, a further analysis is proposed in
section \ref{SecMn_MnInteraction}, 
where
cohesive energies are compared using
realistic TM-TM pair interaction.

\begin{figure}
\begin{center}
\psfig{file=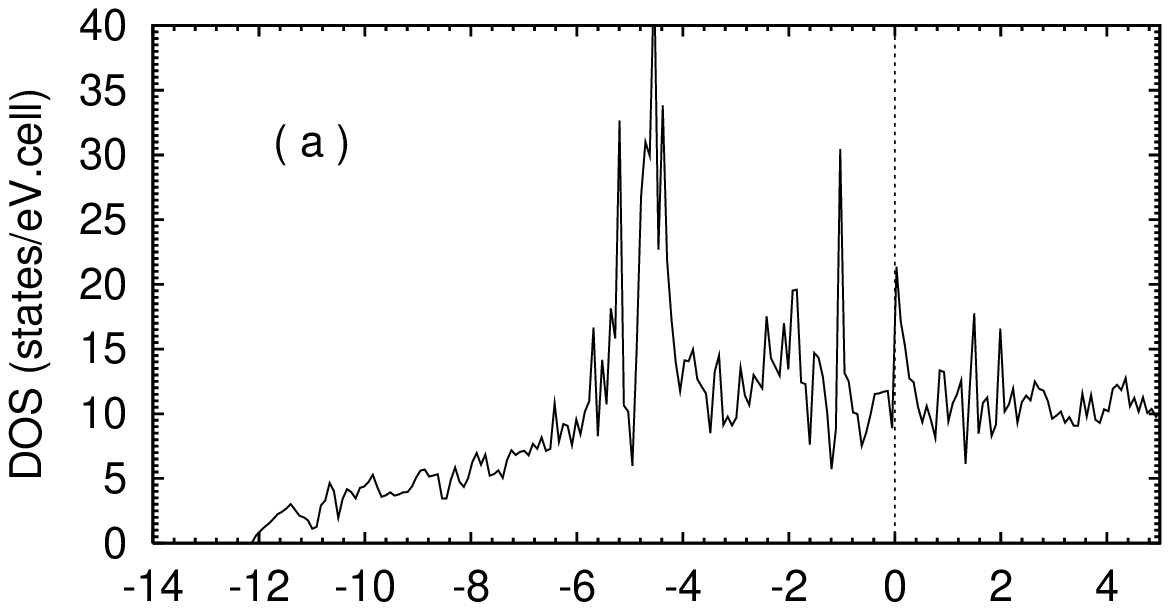,width=7cm}
\hspace{-0.85cm}
\psfig{file=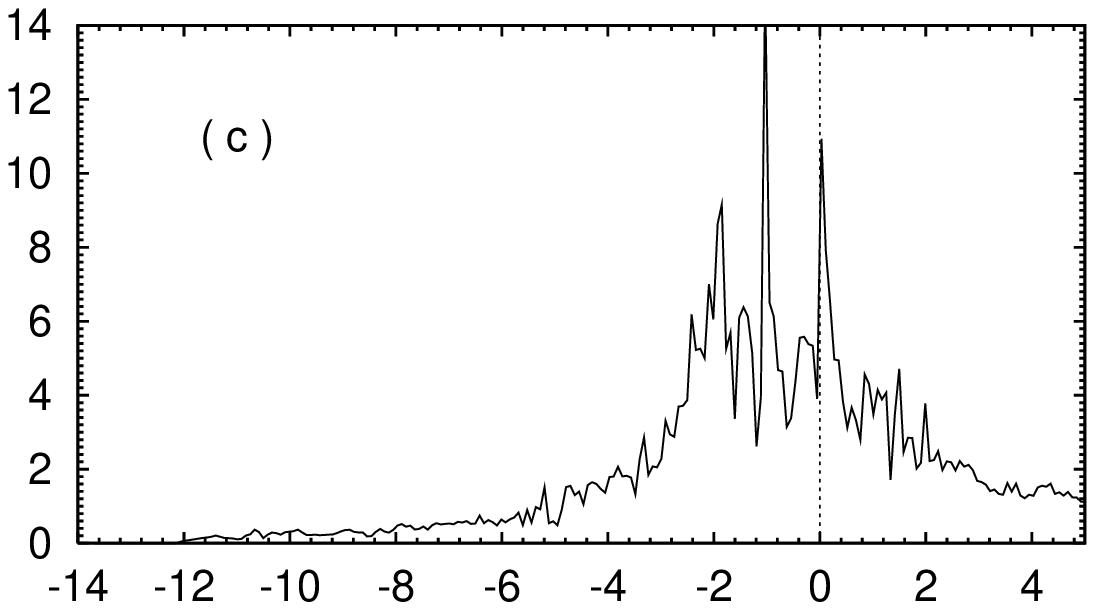,width=7cm}

\vspace{-0.5cm}
\psfig{file=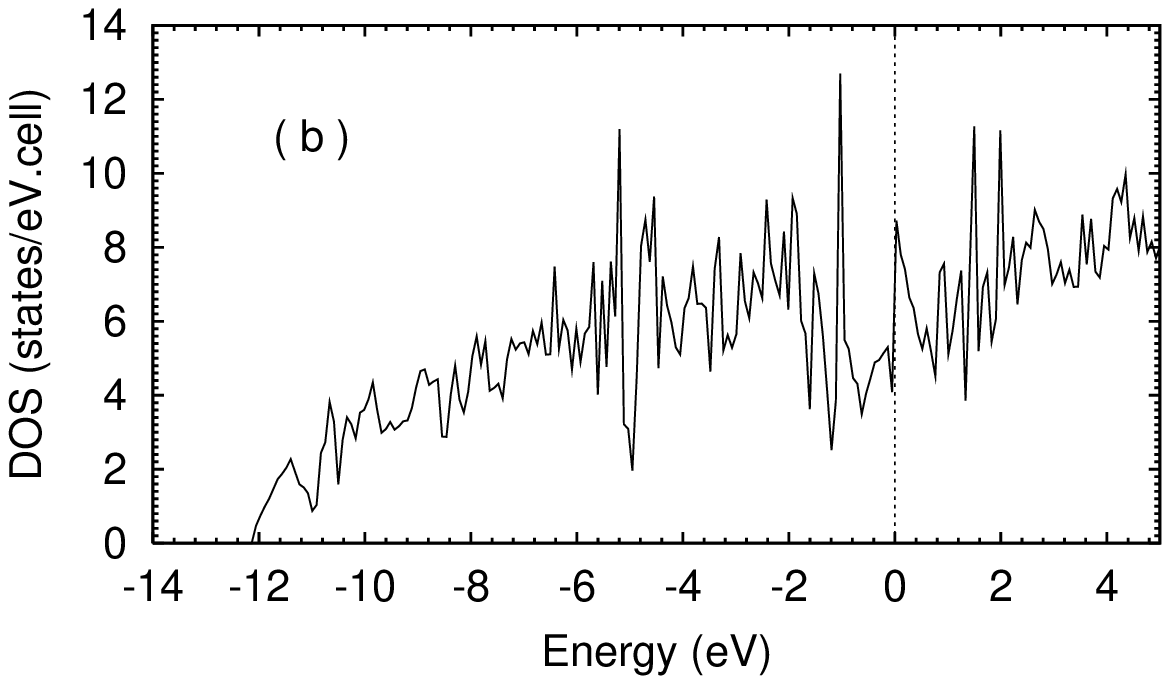,width=7cm}
\hspace{-0.85cm}
\psfig{file=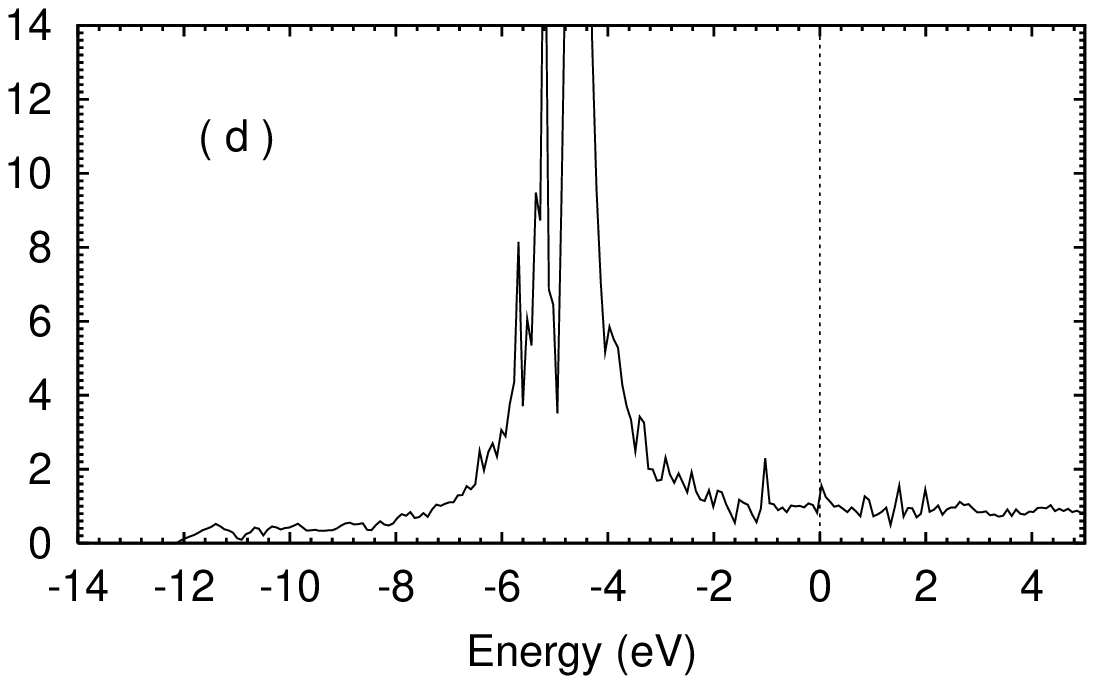,width=7cm}

\caption{LMTO-ASA DOSs performed
by LMTO-ASA method in hypothetical
$\rm \beta\,Al_9Mn_{1.5}Cu_{1.5}Si$:
(a) total DOS,
(b) \{Al\,+\,Si\} local DOS,
(c) Mn local DOS,
(d) Cu local DOS.
The hypothetical
$\rm \beta\,Al_9Mn_{1.5}Cu_{1.5}Si$ is
built by placing 1 Mn-triplet by 1 Cu-triplet
in each unit cell of $\rm \beta\,Al_9Mn_{3}Si$.
$E_F=0$.}
\label{FigDOSAlMnSiCu}
\end{center}
\end{figure}

\subsubsection*{(iii) Effect of Mn-Mn medium range interaction\\ \\}
\label{Sec_B_AlMnCuSi}

\begin{table}
\caption{\label{Tab_DistanceMn-Mn} Mn-Mn distances in
$\rm \beta\,Al_9Mn_3Si$ and hypothetical
$\rm \beta\,Al_9Mn_{1.5}Cu_{1.5}Si$ (see text).}
\begin{indented}
\item[]\begin{tabular}{@{}lll}
\br
Mn-Mn distance & \multicolumn{2}{l}
{Number of Mn-Mn pairs} \\
($\rm \AA$) & $\rm \beta\,Al_9Mn_3Si$ &
$\rm \beta\,Al_9Mn_{1.5}Cu_{1.5}Si$ \\
\mr
2.69 & 2 & 2 \\
4.17 & 4 &   \\
4.83 & 2 & 2 \\
4.96 & 2 &  \\
6.38 & 4 & \\
6.59 & 4 & 4 \\
7.33 & 8 &  \\
7.51 & 6 & 6 \\
\br
\end{tabular}
\end{indented}
\end{table}

The Mn-Mn distances in $\rm \beta\,Al_9Mn_3Si$ are reported 
in table
\ref{Tab_DistanceMn-Mn}.
Mn are grouped together to form Mn-triplets
(section \ref{SecStructure_General}). 
In order to determine the effect of a possible
Mn-Mn medium range interaction
on a pseudogap, a LMTO calculation was performed on a
modified $\beta$ phase containing only one Mn-triplet per
unit cell instead of two. 
In this case,  $\rm
\beta\,Al_9Mn_3Si$ was transformed into
$\rm \beta\,Al_9Mn_{1.5}Cu_{1.5}Si$ 
by remplacing a Mn-triplet by a Cu-triplet.
Mn environments remain identical  up to $\rm 4.17\,\AA$
(table \ref{Tab_DistanceMn-Mn}).

As results,
the local Cu DOS (mainly d states, 
figure \ref{FigDOSAlMnSiCu}(d)) 
at $E_F$ is very small;
Cu having almost
the same number of sp electrons as Mn, 
it has a minor effect near $E_F$.
For the local Mn DOS 
the pseudogap disappears completely
(figure \ref{FigDOSAlMnSiCu}(c)).
For the total DOS, 
a small depletion below $E_F$ is still remaining
(figure \ref{FigDOSAlMnSiCu}(a)),
and for the local \{Al\,+\,Si\} DOS 
there is a pseudogap below $E_F$ 
(figure \ref{FigDOSAlMnSiCu}(b)), 
but
less pronounced than for
$\rm \beta\,Al_9Mn_3Si$ (figure \ref{FigDOSAlSi}(a)).
Therefore, such a disappearence of  pseudogap
proves the  effect
of Mn-Mn interactions over medium distances equal to
4.17, 4.96, 6.38 $\rm \,\AA$\ldots
(table \ref{Tab_DistanceMn-Mn}).

%
%

\section{Effective Bragg potential for sp states}
\label{BraggPotential}

\subsection{Exitence of  effective sp hamiltonian}

As electrons are nearly-free electrons 
in Hume-Rothery sp crystals without TM atoms
the hamiltonian writes \cite{Massalski78},
\begin{eqnarray}
H_{sp}=\frac{\hbar^2\,k^2}{2m}+V_B.
\label{Hamil_sp}
\end{eqnarray}
$V_B$ is a weak potential 
(Bragg potential), and does not depend on 
an energy,
\begin{eqnarray}
V_B({\bf r})=\sum_{\bf K}V_B({\bf K})\,{\rm e}^{i{\bf K}\cdot{\bf r}},
\label{V_Bragg_sp}
\end{eqnarray}
where  the vectors ${\bf K}$ belong to the reciprocal lattice.
However, for alloys containing TM atoms,
the strong scattering of sp electrons by TM atoms can not
be described from a weak potential.
In this case, a
generalised Friedel-Anderson hamiltonian \cite{Anderson61}
has been considered.
In a non-magnetic case:
\begin{eqnarray}
H = H_{sp} + H_d + H_{sp\textrm{-}d}\;,
\label{H_Friedel_Anderson}
\end{eqnarray}
where sp states are delocalised nearly-free states 
(equation (\ref{Hamil_sp})) 
and d states are localised on d orbitals of 
TM atoms. $H_d$ is the energy of d states. 
The term $H_{sp\textrm{-}d}$
represents a sp-d coupling which is essential in
this context.
The eigenstates $\psi$ of $H$ can be decomposed in two terms:
\begin{eqnarray}
|\Psi \rangle = |\Psi_{sp}  \rangle + |\Psi_{d} \rangle\;,
\end{eqnarray}
where $\Psi_{sp}$ and $\Psi_{d}$ are each linear combinations of
sp states and linear combinations of
d orbitals of all TM atoms.
The classical tight-binding approximation
$\langle \Psi_{sp} | \Psi_{d} \rangle=0$ is made.

An
{\it ``effective Bragg potential''} for sp states,
including effects of d orbitals of TM
atoms,
is calculated in order to analyse the effect of TM atoms.
A projection of the Schr\"odinger
equation, $(H-E)|\Psi \rangle=0$, on the
sub-space generated by sp states allows one to
write the effective hamiltonian for sp states: 
\begin{eqnarray}
H_{eff(sp)}= \frac{\hbar^2\,k^2}{2m} + V_{B,eff}
~~{\rm with}~~V_{B,eff}=
V_B + H_{sp\textrm{-}d}\,
\frac{1}{E-H_d}\, H_{sp\textrm{-}d}\,,
\label{Hamil_eff_sp}
\end{eqnarray}
and where
$V_B$ is as given by equation 
(\ref{V_Bragg_sp}).
The second term of $V_{B,eff}$ depends on energy.
In crystals and quasicrystals, 
$V_{B,eff}$ is an
effective Bragg potential
that takes into account the scattering of sp states by 
the strong potential of TM atoms.

\subsection{Characteristic of effective Bragg potential}
\label{BraggPotential2}

For the phases presently considered, 
there are a few 
pairs of Mn atoms that are near-neighbours. Indeed each 
Mn is surrounded by 10 Al (Si) and 2 Mn 
(section \ref{SecStructure}). 
Therefore, a direct hoping between 
two d orbitals can be neglected. Thus:
\begin{eqnarray}
 H_d = \sum_{d,i} E_{di} \, | d,i\rangle
\langle d,i |\;,
\end{eqnarray}
where $i$ is a TM site index and $d$ the 
five d orbitals
of each TM atom.
Assuming that all TM atoms are equivalent,
one has $E_{di}=E_d$.
The Fourier coefficients of the effective Bragg potential
$V_{B,eff}$ are  calculated
from $H_{eff(sp)}$ using the formula
$V_{B,eff}({\bf K})=\langle {\bf k}|H_{eff(sp)}
|{\bf k}-{\bf K} \rangle$. One obtains:
\begin{eqnarray}
V_{B,eff}({\bf r}) = \sum_{\bf K} 
\left( V_B({\bf K}) + \frac{
|t_{{\bf k},{\bf K}}|^2}{E - E_d} 
\sum_i e^{-i {\bf K}.{\bf r}_{i}} \right)
e^{i{\bf K}.{\bf r}}\;, \label{EqVeffectif}\\
{\rm where}~~~|t_{{\bf k},{\bf K}}|^2 = 
\sum_{d=1}^5
\langle {\bf k}|
H_{sp\textrm{-}d} |d_0\rangle
\langle d_0 | H_{sp\textrm{-}d} |
{\bf k}-{\bf K} \rangle\;,
\end{eqnarray}
where 
${\bf r}_i$ is the position of TM(i) atoms. 
By convention a TM
atom with orbital $d_0$ is on a site 
at ${\bf r}_0=0$.
$t_{{\bf k},{\bf K}}$ is a matrix element that
couples sp states $|{\bf k}\rangle$ and
$|{\bf k}-{\bf K}\rangle$ via  sp-d hybridisation.
The expression
(\ref{EqVeffectif}) is exact providing that a direct d-d
coupling is neglected.

The potential of TM atoms is strong and creates d resonance
of the wave function in an energy range
$ E_d-\Gamma \leq E \leq E_d+\Gamma$,
where
$2\Gamma$ is the width of the d resonance.
In this energy range,
the second term of equation (\ref{EqVeffectif}) is
essential 
as it does represent
the diffraction of the sp electrons by a network
of d orbitals,
i.e. the  factor
$\left(\sum_i e^{-i {\bf K}.{\bf r}_{di}}\right)$
corresponding to the structure factor of the
TM atoms sub-lattice.
As the d band of Mn is almost half filled,
$E_F \simeq E_d$, this factor is important
for energy close to $E_F$.
Note that 
the Bragg planes associated with the second term of
equation (\ref{EqVeffectif}) correspond to
Bragg planes determined by diffraction.
For $\beta$ phase, it can be concluded that $V_B$ has no 
effect on a pseudogap and on a phase stabilisation
because of absences of pseudogap for DOS
calculated for $\rm \beta\,Al_9Mn_3Si$
without sp-d hybridisation (figure~\ref{FigDOSAlSi}(b))
and
$\rm \beta\,Al_9Al_3Si$ without 
Mn atoms (figure~\ref{FigDOSAlSi}(c)).
Let us note
however that the Hume-Rothery mechanism for alloying still 
minimizes the sp band energy due to
a strong scattering of sp states by the Mn 
sub-lattice.

In summary, an analyse in term of effective Bragg potential
allows one to interpret  LMTO results 
as hybridisation-induced pseudogap 
in total and sp DOSs 
which comes from a
diffraction
of sp states by the sub-lattice of Mn atoms via the sp-d
hybridisation.
In this context the medium range distance between TM 
atoms might have important role.

%
%
\section{Role of indirect Mn-Mn pair interaction}
\label{SecMn_MnInteraction}

\subsection{Medium range  TM-TM interaction in Al based alloys}

\begin{figure}
\begin{center}
\psfig{file=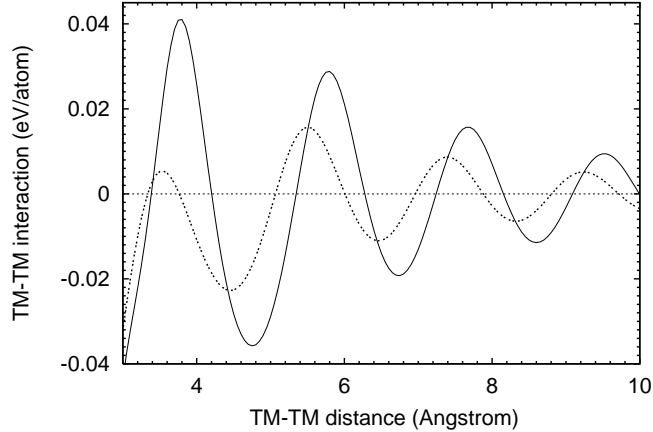,width=9cm}
\caption{Indirect (solid lines) Mn-Mn  pair
interaction $\Phi_{Mn\textrm{-}Mn}$ and (dashed line) Co-Co
pair interaction $\Phi_{Co\textrm{-}Co}$.
This interaction does not include the short range
repulsive term between two TM atoms. 
TM atoms are non-magnetic.}
\label{FigPotentiel}
\end{center}
\end{figure}

As
a Hume-Rothery stabilisation is a consequence of
oscillations of charge density of valence electrons with
energy close to $E_F$
\cite{Blandin,Massalski78,Hafner79,Hafner87,Pettifor00}, 
a most stable atomic structure
is obtained when distances between atoms are multiples of
the wavelength $\lambda_F$ of  electrons with energy
close to $E_F$.
Since
the scattering of  valence sp states by the
Mn sub-lattice is strong,
the Friedel oscillations of charge of sp electrons
around Mn  must have a strong effect on a stabilisation.
Taking into account that  
stabilisation occurs for a 
specific Mn-Mn distance of 
4.7\,$\rm \AA$ \cite{ZouPRL93}.
A Hume-Rothery mechanism in Al(rich)-TM compounds
might be analysed in term of
an indirect medium range TM-TM pair interaction
resulting from a strong sp-d hybridisation.
Zou and Carlsson \cite{ZouPRL93,Zou94} have 
calculated this interaction
from an Anderson
model hamiltonian with two impurities, using a Green's function
method.
A calculation of an indirect TM-TM pair interaction, 
$\Phi_{TM\textrm{-}TM}$,
within a multiple scattering approach \cite{GuyPRB97}
yields a result in good
agreement with this given in Ref. \cite{ZouPRL93}
(figure~\ref{FigPotentiel}).
Mn-Mn \cite{ZouPRL93,Zou94} and
Co-Co \cite{Phillips94,Moriarty97}  interactions
have been used successfully for
molecular-dynamics studies
\cite{Mihalkovic96,Mihalkovic01} of Al-Mn and Al-Co 
systems near the
composition of quasicrystals.

As interaction 
magnitudes 
are larger for TM-TM than
for Al-TM and Al-Al, 
$\Phi_{TM\textrm{-}TM}$
has a major effect on the electronic
energy.
Because of the sharp Fermi surface
of Al, it 
asymptotic form at large TM-TM distance ($r$) is of the form:
\begin{eqnarray}
\Phi_{TM\textrm{-}TM}(r) \propto \frac{\cos (2k_F\,r-\delta)}{r^3}\;.
\label{EqVeffOscil}
\end{eqnarray}
The phase shift $\delta$ 
depends on the nature of the TM atom and varies from $2\pi$
to $0$ as the d band fills.
Magnitude of the medium range interaction is larger 
for Mn-Mn than
for other transition metal (Cr, Fe, Co, Ni,
Cu), because the number of  d electrons 
close to $E_F$ is the largest for Mn, 
and
the most delocalised electrons are electrons with
Fermi energy.
From the figure~\ref{FigPotentiel}
and equation (\ref{EqVeffOscil}),
it is clear  that distances corresponding to
minima of $\Phi_{TM\textrm{-}TM}$ 
depend also on the nature of TM atom.

\subsection{Contribution of the medium range Mn-Mn 
interaction to total energy}
\label{Sec_Mn-Mn_Energy}

The {\it ``structural energy''}, $\cal{E}$, of TM sub-lattice in
Al(Si) host is defined as the energy needed to built the
TM sub-lattice in the metallic host that simulates 
Al and Si atoms 
from isolated TM atoms in the metallic host.
$\cal{E}$ per unit cell is:
\begin{eqnarray}
{\cal E} =  \sum_{i,j\,(j\neq j)} 
\frac{1}{2} \,
\Phi_{TM\textrm{-}TM}(r_{ij})~e^{-\frac{r_{ij}}{L}}\;,
\label{EquationEStruturale}
\end{eqnarray}
where $i$ and $j$ are index of TM atom and
$r_{ij}$,  TM(i)-TM(j) distances.
$L$ is the mean-free path of electrons due to scattering
by static disorder or phonons \cite{LibreParMoyen}.
$L$  depends on the structural quality
and temperature and can only be estimated to be larger
than 10\,$\rm \AA$.
Note that a similar exponential damping factor was introduced 
originaly in the
treatment of RKKY interaction \cite{deGennes62,Blandin}.
In the following, the effects 
of TM-TM pairs over distances larger
than first-neighbour distances is analysed.
Therefore, an energy $\cal{E}'$ is calculated
from equation (\ref{EquationEStruturale}) without including
first-neighbour TM-TM terms in the sum.
$\cal{E}'$ is the part of the structural energy of TM sub-lattice
that only comes from medium range distances.

\begin{figure}
\begin{center}
\psfig{file=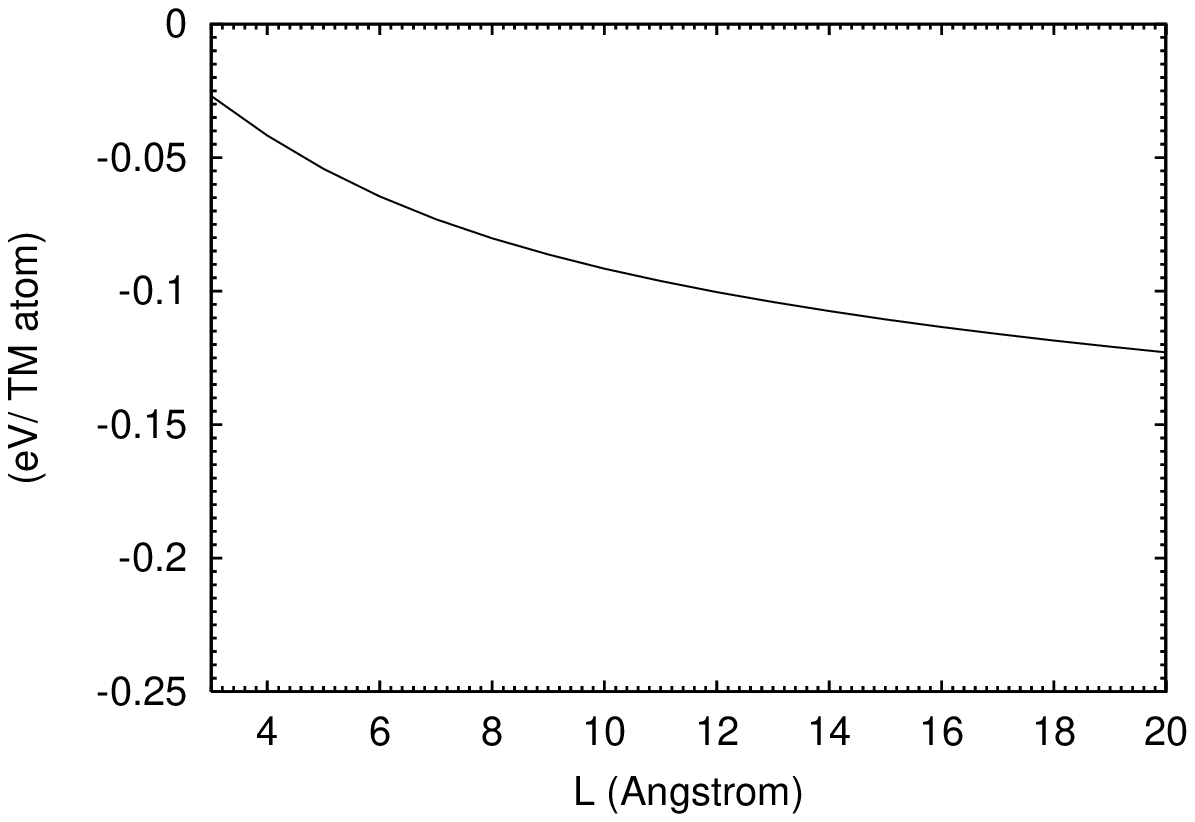,width=9cm}
\caption{Structural energy ${\cal E}'$ of the Mn sub-lattice
in (line) $\rm \beta\,Al_9Mn_3Si$,
({\scriptsize $\triangle$}) $\rm \varphi\,Al_{10}Mn_3$,
($\bullet$) $\rm o\,Al_6Mn$,
($\times$) $\rm \alpha\,$Al-Mn-Si,
and ($\diamond$) hypothetical
$\rm \beta\,Al_9Mn_{1.5}Cu_{1.5}Si_{1.5}$.}
\label{FigEBetaPhiAlphaAl6Mn}
\end{center}
\end{figure}

Structural energies, ${\cal E}'$,
of the Mn sub-lattice are shown for  
$\rm \beta\,Al_9Mn_3Si$ and 
$\rm \varphi \,Al_{10}Mn_3$ 
structures in figure \ref{FigEBetaPhiAlphaAl6Mn}, 
where 
they are compared to those of 
$\rm o\,Al_6Mn$ \cite{Villars85}  and
$\rm \alpha$\,Al-Mn-Si approximants
\cite{Guyot85,ElserH85}.
${\cal E}'$ are always negative
with
magnitudes less than $-0.1$\,eV\,$/$\,TM atom,
but 
strong enough to give a significant contribution
to the band energy.

This result is in good agreement 
with an effect of
Mn sub-lattice on the pseudogap as shown 
previously (sections \ref{Sec_spd}
and \ref{BraggPotential}).
According to a Hume-Rothery mechanism, one expects that 
a pseudogap is well pronounced
for a large value of $|{\cal E}'|$.
Such a  correlation is verified for the hypothetical
$\rm  \beta\,Al_9Mn_{1.5}Cu_{1.5}Si$  
(section \ref{Sec_B_AlMnCuSi}) where 
the  diminution of pseudogap 
in $\rm  \beta\,Al_9Mn_{1.5}Cu_{1.5}Si$
sp DOS (figure \ref{FigDOSAlMnSiCu}(b)) 
with respect 
to $\rm \beta\,Al_9Mn_{3}Si$ 
sp DOS (figure \ref{FigDOSAlSi}(a)), 
corresponds to  reduction
of $(|{\cal E}'|)$ 
(figure \ref{FigEBetaPhiAlphaAl6Mn}).

\subsection{Origin of the Vacancy}

For structures containing serveral Mn  Wyckoff sites,
the TM-TM pair interaction mediated by conduction states allows 
one
to compare the relative stability of TM atoms on
different Wyckoff sites.
Considering a phase with a structural energy  
of the TM sub-lattice equal to ${\cal E}$,
the variation, $\Delta {\cal E}_i$, of ${\cal E}$
is determined
when 
one TM(i) atom is removed from the structure:
\begin{eqnarray}
\Delta {\cal E}_{i}=-\sum_{j\,(j\neq i)}
\Phi_{TM\textrm{-}TM}(r_{ij})~e^{-\frac{r_{ij}}{L}}\;.
\label{EquationSumDE}
\end{eqnarray}
TM atoms on different Wyckoff sites 
have different $\Delta {\cal E}_i$ values that can 
be compared.
The most stable Mn sites correspond to highest 
$\Delta {\cal E}_i$ values.
Moreover, the energy reference is a TM
imputity in the Al(Si) matrix which 
does not depend on the structure. 
Therefore, 
it is possible to compare $\Delta {\cal E}_i$
calculated for different structures. 
As previously, the 
energy $\Delta {\cal E}_i'$ is calculated from equation 
(\ref{EquationSumDE}) without the first-neighbour 
TM-TM contributions in order to analyse 
effects at medium range order.

\begin{figure}
\begin{center}
\psfig{file=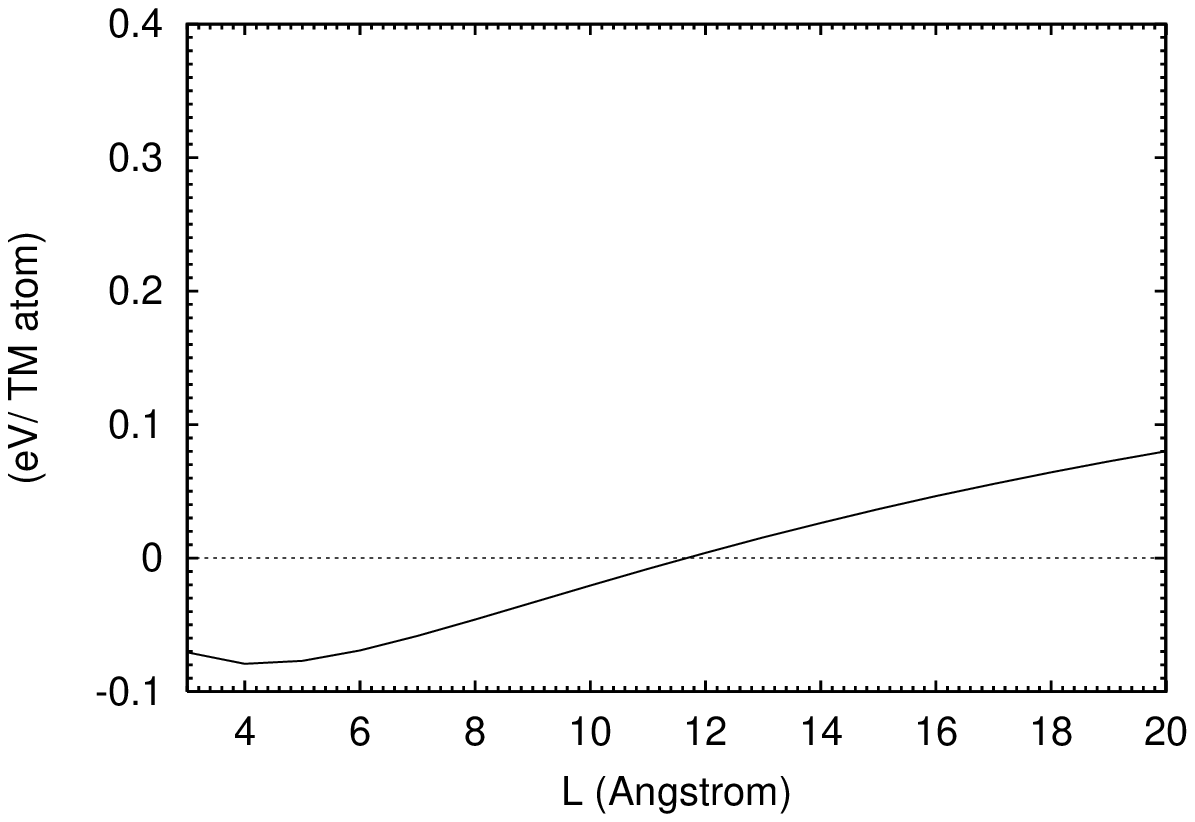,width=9cm}
\caption{Variation of the structural energy 
$\Delta {\cal E}_i'$ due to
Mn-Mn interaction  for (simple line) Mn(0) 
in   (2d) in the hypothetical
$\rm \beta\,Al_9Mn_4Si$;
for
({\scriptsize $\triangle$}) Mn in (6h) in the hypothetical
$\rm \beta\,Al_9Mn_4Si$;
for
($\bullet$) Mn(1) in (2b) in
$\rm \mu\,Al_{4.12}Mn$ \cite{Shoemaker89};
and for ($\times$) Mn(1) in (2d) in
$\rm \lambda\,Al_4Mn$ \cite{Kreiner97}.
Mn(1) in
$\rm \mu\,Al_{4.12}Mn$ and $\rm \lambda\,Al_4Mn$
have similar local environment as 
Mn(0) in hypothetical
$\rm \beta\,Al_9Mn_4Si$.}
\label{FigEVaMn1}
\end{center}
\end{figure}

\begin{figure}
\begin{center}
\psfig{file=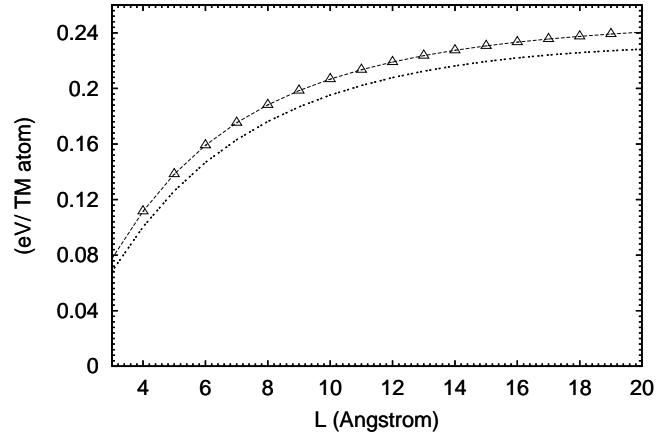,width=9cm}
\caption{Variation of the structural energy
$\Delta {\cal E}_i'$ due to Co-Co 
interaction in $\rm Al_5Co_2$.
$\Delta {\cal E}_i'$  is calculated for the 
two Co Wyckoff sites:
(simple dashed line) Co(0) in (2d);
({\scriptsize $\triangle$}) Co(1) in (6h).}
\label{FigEAl5Co2}
\end{center}
\end{figure}

Considering the hypothetical 
$\rm \beta\,Al_9Mn_4Si$
builted from $\rm \beta\,Al_9Mn_3Si$ 
in which a Mn atom (Mn(0)) replaces a
vacancy (Va) in (2d), 
it appears that 
$\Delta {\cal E}_{Mn(1)}' > \Delta {\cal E}_{Mn(0)}'$
(figure \ref{FigEVaMn1}).
Mn(0) in (2d)
is therefore less stable than Mn in (6h) for hypothetical 
$\rm \beta\,Al_9Mn_4Si$,
thus justifying that a vacancy exists in $\beta$ phase.
A similar result was obtained for 
$\rm \varphi\,Al_{10}Mn_3$.

On opposite,
for complex crystals 
$\rm \mu\,Al_{4.12}Mn$ \cite{Shoemaker89}
and $\rm \lambda\,Al_4Mn$ \cite{Kreiner97}
containing a Mn site 
(Mn(1) in Refs \cite{Shoemaker89,Kreiner97})
with similar
local environment as Va (or Mn(0)) in 
$\beta$ structure (section \ref{SecStructure_Vancancy}),
the corresponding
$\Delta {\cal E}_i'$ values differs strongly from those of
Mn(0) in hypothetical $\rm \beta\,Al_9Mn_4Si$.
Thus Mn(1) in $\mu$
and $\lambda$ are more stable 
than an additional Mn atom replacing the vacancy
in  $\rm \beta$ and 
$\rm \varphi$.
Moreover, both
$\Delta {\cal E}_{Mn(1)}'$ in $\mu$ and
$\lambda$ have the same order of magnitude as
the $\Delta {\cal E}_i'$  calculated for
other Mn(i) atoms in $\mu$ and $\lambda$ 
($\mu$ 
and $\lambda$ phases
contain 10 and 15 Mn Wyckoff sites
respectively \cite{Shoemaker89,Kreiner97}). 
Thus Mn(1) in $\mu$ and
$\lambda$ is stable.
Such a  difference
between $\beta$, $\varphi$ and 
$\mu$, $\lambda$ 
can be interpreted in terms 
of medium range
Mn-Mn distances
with respect to the curve of 
figure \ref{FigPotentiel}:
In
$\beta$, $\varphi$ phases,
environment of Va  contains two Mn at distance
{3.8\,\AA} (table~\ref{Tab_Distances}),
whereas the smallest Mn(1)-Mn distance is
{4.8\,\AA} in $\mu$ and $\lambda$ phases.
{3.8\,\AA}
corresponds to an unstable Mn-Mn distance whereas 
{4.8\,\AA} corresponds to a stable one
(figure \ref{FigPotentiel}).

For
$\rm Al_5Co_2$ phase almost isomorphic of
$\beta$ and $\varphi$ phases,
there is a Co site (Co(0)) corresponding to the vacancy
of $\beta$ and $\varphi$ (table \ref{Tab_Structure}).
In this case $\Delta {\cal E}_{Co(0)}'$, calculated
with a Co-Co pair interaction, is almost equal
to $\Delta {\cal E}_{Co(1)}'$ (figure~\ref{FigEAl5Co2}).
As Co(0) in (2d) is  as stable as Co(1) in (6h), 
it justifies why
any vacancy does not exist in $\rm Al_5Co_2$.

The present anaylsis on the origin of the vacancy 
in terms of TM-TM medium range interactions confirms 
the LMTO results (section~\ref{LMTO_Vacancy}). 
It shows the importance
of TM-TM medium range indirect interaction
on the atomic structure.

\section{ Magnetic properties }
\label{SecMagnetism}

The presence of localised magnetic moments in quasicrystals and related
phases containing Mn is much debated
\cite{GuyEuroPhys93,Hafner98a,art_prejean,
Bratkovsky95,Bahadur97,Hippert_Liquide,Virginie2,
Krajci98,Simonet98,Hippert99,Prejean02}.
Vacancies, Mn pairs, triplets, quadruplets,
quintuplets, variation of first-neighbour distances around Mn
are often invoked to
explain magnetic moments
\cite{Cooper76,Hoshino93,Guenzburger91,Guenzburger94,
Scheffer00,Hafner98a,Hippert99}.
But in previous work \cite{GuyPRL00,GuyICQ7},
it has been shown that an extreme
sensitivity of magnetic properties also comes
from an effect of an indirect Mn-Mn
interaction mediated
by sp states.
Consequently an analysis limited to first-neighbour
environments is not sufficient
to interpret magnetic properties.

The unit
cell of $\beta$ and $\varphi$ phases
contains two Mn-triplets distant each other from 
from about $\rm 5\,\AA$, 
and
experimental
measurements indicate that Mn-triplets are 
non-magnetic \cite{Virginie2}.
LMTO electronic structures calculated with polarised spin
confirms that Mn triplets are non-magnetic in
$\rm \beta\,Al_9Mn_3Si$ and $\rm \varphi\,Al_{10}Mn$.
But from polarised spin  LMTO calculation,
performed on
$\rm \beta\,Al_9Mn_{1.5}Cu_{1.5}Si$ phase
where a Cu-triplet 
replaces one Mn-triplet in
each cell (section \ref{Sec_B_AlMnCuSi}$(iii)$ and table 
\ref{Tab_DistanceMn-Mn}),
a
magnetic moment equal to $\rm 1\,\mu_B$ was found 
on each Mn in $\rm
\beta\,Al_9Mn_{1.5}Cu_{1.5}Si$ 
(i.e.  3 Mn in Mn-triplet are almost
equivalent with a ferromagnetic spin orientation).
The energy of formation of magnetic
moments in $\rm \beta\,Al_9Mn_{1.5}Cu_{1.5}Si$
is $-0.046\,\rm eV$
per triplet.
The Cu has no long range interaction as its d orbitals are
full.
Thus a medium range Mn-Mn interaction
holds Mn-triplets in non-magnetic state whereas
a Mn-triplet impurity in Al should be magnetic.
It proves
that a magnetic state of a Mn atom is very sensitive to 
surrounding
Mn atoms at a medium range distance
up to $\rm 4.17\,\AA$
(table \ref{Tab_DistanceMn-Mn}).
The model of
the spin polarised
Mn-Mn interaction 
presented in Ref. \cite{GuyPRL00} is in agreement
with this LMTO result.

As explain in the literature 
\cite{GuyEuroPhys93,Bratkovsky95,Hafner98a,Krajci98,GuyICQ7},
the occurrence of magnetic
Mn can be related to a reduction of pseudogap
in the local Mn paramagnetic
DOS in $\rm \beta\,Al_9Mn_{1.5}Cu_{1.5}Si$
(figure \ref{FigDOSAlMnSiCu}(c))
by comparison with the local Mn DOS in 
$\rm \beta\,Al_9Mn_{3}Si$ (figure \ref{FigDOSlocalBeta}).
However, the present study shows that a pseudogap in 
paramagnetic Mn DOS
does not only depend  on the local environment of Mn
as it is also very sensitive
to Mn-Mn medium range interaction
(sections \ref{Sec_spd}$(iii)$, 
\ref{BraggPotential2} and \ref{Sec_Mn-Mn_Energy}).

\section{Conclusion}
\label{SecConclusion}

From  first-principles calculations
combined with a
model hamiltonian approach,
it is shown that a detailed analysis of
the electronic structure
allows one to explain the following features of 
$\rm \beta\,Al_9Mn_3Si$ and 
$\rm \varphi\,Al_{10}Mn_3$ structures:
\begin{itemize}
\item The  small amount of Si in $\rm \beta\,Al_9Mn_3Si$
stabilises its structure due to a 
shift of the Fermi energy toward the minimum of the
pseudogap in the DOS.
An
{\it ab initio} study shows that, 
at 0\,Kelvin, Si atoms are on a Wyckoff site 
different of those for Al atoms.
\item $\rm \beta\,Al_9Mn_3Si$ and 
$\rm \varphi\,Al_{10}Mn_3$ are sp-d Hume-Rothery
phases. 
The transition metal (TM) elements
have 
a crucial effect as sp electrons are scattered 
by an effective Bragg potential dominated by
the effect of the Mn sub-lattice. 
Such a Bragg potential relates to an
indirect Mn-Mn interaction which has 
a strong magnitude up to 
$\rm \sim 5\,\AA$ and more.
\item An analysis in terms of medium range
TM-TM interactions gives theoretical arguments
to understand the origin of a large
vacancy existing in $\rm \beta\,Al_9Mn_3Si$
and $\rm \varphi\,Al_{10}Mn_3$, 
whereas similar sites
are occupied by Mn in $\rm \mu\,Al_{4.12}Mn$
and $\rm \lambda\,Al_4Mn$, and by Co
in $\rm Al_5Co_2$.
\end{itemize}

Finally, in $\rm \beta\,Al_9Mn_3Si$ and
$\rm \varphi\,Al_{10}Mn_3$,
the Hume-Rothery minimization of
band energy leads to
a ``frustration'' mechanism which favours a complex
atomic structure.
The Mn sub-lattice appears 
to be the skeleton of the stucture via
a medium range indirect interactions 
between Mn atoms in the Al matrix.
As $\beta$ and $\varphi$ structures
are related to those of quasicrystals, it suggests that
Hume-Rothery stabilisation, expressed in terms of
Mn-Mn interactions,
is intrinsically linked to the emergence of
quasiperiodic structures in Al(Si)-Mn systems.

\section*{Acknowledgments}
I am very grateful to D. Mayou with whom many
ideas of this work have been developed.
I thank also F. Hippert, D. Nguyen-Manh, R. Bellissent,
V. Simonet and J. J. Pr\'ejean 
for fruitful discusions.
My thanks to M. Audier from stimulating discussions
and carefull reading of this paper.

\section*{References}


\end{document}